\documentclass[sigconf, nonacm]{acmart}
\AtBeginDocument{%
  }

\setcopyright{none} 
\acmConference[KDD '26]{ACM SIGKDD Conference on Knowledge Discovery and Data Mining}{August 9--13, 2026}{Jeju, Korea}
\acmYear{2026}
\copyrightyear{2026}

\acmSubmissionID{83}

\usepackage{url}
\usepackage{makecell}
\usepackage{multirow}
\usepackage{pifont}
\usepackage{rotating}
\usepackage{subcaption}
\usepackage{listings}
\usepackage{xcolor} 
\usepackage{stfloats}
\newcommand{\cmark}{\checkmark}
\newcommand{\xmark}{$\times$}

\lstdefinestyle{pythonsnippet}{
  language=Python,
  basicstyle=\ttfamily\small,
  numbers=left,
  numberstyle=\tiny,
  stepnumber=1,
  numbersep=8pt,
  showstringspaces=false,
  breaklines=true,
  tabsize=2,
  frame=single,
  rulecolor=\color{black!20},
  keywordstyle=\bfseries,
  commentstyle=\itshape\color{black!55},
  stringstyle=\color{black}
}

\definecolor{codegray}{rgb}{0.95,0.95,0.95}
\definecolor{framegray}{rgb}{0.8,0.8,0.8}

\lstdefinestyle{promptstyle}{
    backgroundcolor=\color{codegray},   
    basicstyle=\ttfamily\small,      
    breaklines=true,                 
    frame=single,                    
    rulecolor=\color{framegray},       
    frameround=tttt,                 
    captionpos=b,                    
    numbers=left,                    
    numberstyle=\tiny\color{gray}, 
    keywordstyle=\color{blue},       
    commentstyle=\color{green!60!black}, 
    stringstyle=\color{purple},      
}

\begin{document}

\title{MMTS-BENCH: A Comprehensive Benchmark for Time Series Understanding and Reasoning}

\author{Yao Yin}
\authornote{Both authors contributed equally to this research.}
\affiliation{%
  \institution{Tsinghua University}
  \city{Beijing}
  \country{China}}
\email{yiny23@mails.tsinghua.edu.cn}

\author{Zhenyu Xiao}
\authornotemark[1]
\affiliation{%
  \institution{Tsinghua University}
  \city{Beijing}
  \country{China}}
\email{xzy23@mails.tsinghua.edu.cn}

\author{Musheng Li}
\affiliation{%
  \institution{Tsinghua University}
  \city{Beijing}
  \country{China}}
\email{lms21@mails.tsinghua.edu.cn}

\author{Yiwen Liu}
\affiliation{%
 \institution{Tsinghua University}
  \city{Beijing}
  \country{China}}
\email{liu-yw23@mails.tsinghua.edu.cn}

\author{Sutong Nan}
\affiliation{%
  \institution{Tsinghua University}
  \city{Beijing}
  \country{China}}
\email{nst24@mails.tsinghua.edu.cn}

\author{Yiting He}
\affiliation{%
  \institution{Tsinghua University}
  \city{Beijing}
  \country{China}}
\email{he-yt24@mails.tsinghua.edu.cn}

\author{Ruiqi Wang}
\affiliation{%
  \institution{Tsinghua University}
  \city{Beijing}
  \country{China}}
\email{wang-rq23@mails.tsinghua.edu.cn}

\author{Zhenwei Zhang}
\affiliation{%
  \institution{Tsinghua University}
  \city{Beijing}
  \country{China}}
\email{thuzhangzw@outlook.com}

\author{Qingmin Liao}
\affiliation{%
  \institution{Tsinghua University}
  \city{Beijing}
  \country{China}}
\email{liaoqm@tsinghua.edu.cn}

\author{Yuantao Gu}
\authornote{Corresponding author}
\affiliation{%
  \institution{Tsinghua University}
  \city{Beijing}
  \country{China}}
\email{gyt@tsinghua.edu.cn}
\renewcommand{\shortauthors}{Yin, Xiao, et al.}

\begin{abstract}
Time series data are central to domains such as finance, healthcare, and cloud computing, yet existing benchmarks for evaluating various large language models (LLMs) on temporal tasks remain scattered and unsystematic. To bridge this gap, we introduce \textsc{MMTS-Bench}, a comprehensive multimodal benchmark built upon a hierarchical taxonomy of time series tasks, spanning structural awareness, feature analysis, temporal reasoning, sequence matching, and cross-modal alignment. \textsc{MMTS-Bench} comprises 2,424 time series question answering (TSQA) pairs across 4 subsets: \textbf{Base}, \textbf{InWild}, \textbf{Match}, and \textbf{Align}, generated through a progressive real-world QA framework and modular synthetic data construction. We conduct extensive evaluations on closed-source, open-source LLMs, and existing time series adapted large language models (TS-LLMs), 
revealing that: (1) TS-LLMs significantly lag behind general-purpose LLMs in cross-domain generalization, (2) LLMs show weaknesses in local tasks compared to global tasks, (3) chain-of-thought (CoT) reasoning and multimodal integration substantially improve performance, and (4) the dominant factor in existing TS-LLMs remains the backbone network capability rather than the time series encoder design. \textsc{MMTS-Bench} not only provides a rigorous evaluation framework but also offers clear directions for advancing LLMs toward robust, interpretable, and generalizable time series reasoning.\footnote{Code and data are available at \url{https://anonymous.4open.science/r/MMTS-BENCH-BEF7/}}
\end{abstract}








\maketitle

\section{Introduction}

Time series data underpin critical systems in finance, healthcare, transportation, and cloud computing \citep{zeng2023financial, haoyietal-informer-2021, liu2024timemmd}, capturing how processes evolve over time. Traditionally, tasks such as forecasting, classification, anomaly detection, and imputation \citep{Yuqietal-2023-PatchTST, zhang2020tapnet, tadnet} rely on specialized statistical models and tooling, demanding substantial domain expertise. In recent years, with the rapid advancement of natural language processing (NLP), especially the breakthroughs in Large Language Models (LLMs) \citep{openai2023gpt4_research,comanici2025gemini,anthropic2025claude37sonnet, qwen2.5, team2025minicpm4}, new possibilities have emerged to overcome the professional barriers in time series analysis \citep{xie2024chatts, wang2025itformer, wang2025chattime, jin2024position}. Integrating time series data with LLMs to build end-to-end time series models has become a prominent research direction \citep{liu2023llava, Qwen2.5-VL}. Recently, a growing number of researchers have begun to explore the application of LLMs to time series analysis, giving rise to novel tasks such as time series description \citep{zhang2023insight}, text-context-assisted forecasting \citep{jin2023time}, simple time series question answering \citep{wang2025chattime}, complex time series reasoning, and cross-variable question answering \citep{xie2024chatts}.

The above works that combine LLMs with time series data also require substantial training and testing datasets for support. Prior efforts either enrich traditional time series datasets with textual annotations \citep{liu2024timemmd, yu2024ecgesi} or build question answering (QA) datasets for time series reasoning \citep{wang2025itformer, kong2025time}. However, most current studies rely on a flat task taxonomy\footnote{Here, \emph{flat} indicates treating tasks such as forecasting, anomaly detection, feature analysis, and high-level temporal reasoning as parallel and independent, without explicit hierarchical relationships.} \citep{wang2025chattime, cai2024timeseriesexam} to define capabilities and synthesize QA data. Such taxonomies either have no hierarchical structure or are simple, making it difficult to comprehensively evaluate LLMs’ abilities in time series understanding and reasoning at a fine-grained level. Moreover, the construction of most time series datasets and the fine-tuning of LLMs are limited to single or small-domain data \citep{wang2025itformer, dong2024fnspidcomprehensivefinancialnews}, and there is still a lack of a comprehensive benchmark across multiple domains to evaluate LLMs’ out-of-distribution (OOD) generalization. More broadly, despite rapid progress, it remains unclear which model design choices are truly critical for effective temporal reasoning and cross-modal alignment, underscoring the need for systematic evaluation frameworks.


\begin{figure}[t]
\begin{center}
\includegraphics[width=\columnwidth]{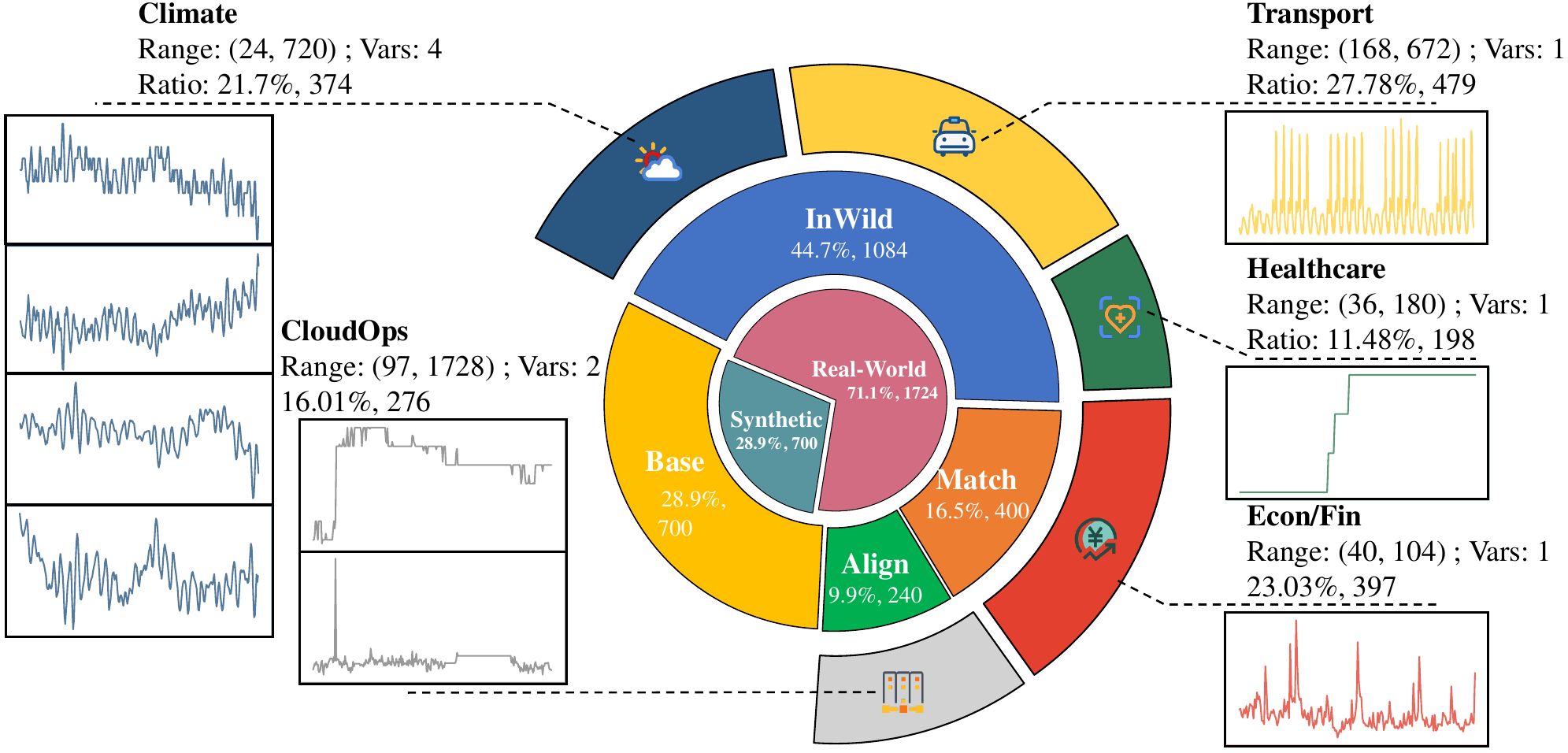}
\end{center}
\caption{MMTS-BENCH overview. Composition of the benchmark across subsets and source domains, with the number of QA instances in each component.}
\label{fig:mmts-overview}
\end{figure}


Inspired by hierarchical capability modeling in multimodal benchmarks \cite{liu2024mmbench} and compositional generalization theory \cite{hupkes2020compositionality}, we propose a hierarchical taxonomy of time series tasks that reorganizes analytical tasks into a multi-level, compositional structure, spanning from basic perception to advanced reasoning, where complex capabilities are formed by systematically combining lower-level task primitives. Compared to prior flat task taxonomies, our hierarchy explicitly models the dependencies and progressive integration among tasks, thereby clarifying and extending several key task areas that have previously been overlooked.

Building on this taxonomy, we construct \textsc{MMTS-Bench}, a new multi-modal, multi-dimensional evaluation benchmark for time series tasks (see Figure \ref{fig:mmts-overview}), comprising 2,424 TSQA pairs across four subsets. One subset is built from synthetic time series data: \textbf{(1) Base}, which assesses capabilities in structural awareness and feature analysis. The others are built from real-world time series spanning five domains (e.g., Transport; see Appendix \ref{sec:real-world sources}) in the LOTSA dataset \citep{woo2024moirai}: \textbf{(2) InWild}, which targets feature analysis and temporal reasoning; \textbf{(3) Match}, which evaluates sequence-similarity matching and morphological correspondence; and \textbf{(4) Align}, which measures bidirectional conversion between time series and natural language as well as advanced cross-modal semantic understanding. We further conduct a comprehensive evaluation of multiple mainstream LLMs using MMTS-Bench, offering fine-grained capability rankings and insights into current strengths and limitations, thereby informing the development of future time series foundation models and benchmarks. Our main contributions are as follows:
\begin{enumerate}
\item \textbf{\textsc{MMTS-Bench}.} We propose a capability-oriented, hierarchical taxonomy of time series tasks and instantiate it in \textsc{MMTS-Bench}, a multimodal, multi-dimensional benchmark comprising 2,424 QA pairs across four subsets (Base, InWild, Match, Align), enabling fine-grained evaluation of abilities ranging from feature analysis to temporal reasoning and cross-modal alignment.
\item \textbf{Progressive real-world TSQA generation.} We present a three-stage framework for generating high-quality QA pairs from real-world time series data, systematically improving generation reliability and scalability for large-scale benchmark construction.
\item \textbf{Controllable synthetic data pipeline.} We design a controllable synthetic data generation pipeline based on modular composition and templated generation, allowing precise control over data diversity and difficulty for targeted assessment of foundational capabilities.
\end{enumerate}

\section{Related Work}

\subsection{Time Series Adapted LLMs}

Multimodal large language models (MLLMs) have demonstrated strong capabilities in natural language processing and cross-modal reasoning. In the domain of TS-LLMs, several implementation paradigms have recently been explored. Time-MQA \citep{kong2025time} serializes time series as textual inputs and reports early gains on the TSQA task. ChatTime  \citep{wang2025chattime} quantizes continuous values into a finite token space, enabling continuous pretraining within a unified LLM framework. \citet{zhuang2024see} uses GPT-4o \citep{openai2024hello_gpt4o} in a two-stage, coarse-to-fine anomaly-detection pipeline over rendered time series plots, while InsightMiner \citep{zhang2023insight} and FinVis-GPT \citep{wang2023finvis} adapt LLaVA \citep{liu2023llava} for time series description and candlestick-chart analysis, respectively.

However, representing dense time series as text or plots significantly increases sequence length and token costs, often with limited performance gains. Alignment-based approaches address this by retaining a dedicated time series encoder and learning a lightweight projector into the LLM embedding space, enabling efficient time-series–text interaction. Based on this paradigm, \citet{chow2024towards} and ChatTS \citep{xie2024chatts} achieve competitive results across classification, description, QA, and reasoning. Nonetheless, a unified benchmark for systematically evaluating the multi-dimensional capabilities of TS-LLMs remains lacking.

\subsection{Time Series QA Datasets}

\begin{figure*}[htbp]
\centering
\includegraphics[width=0.8\textwidth]{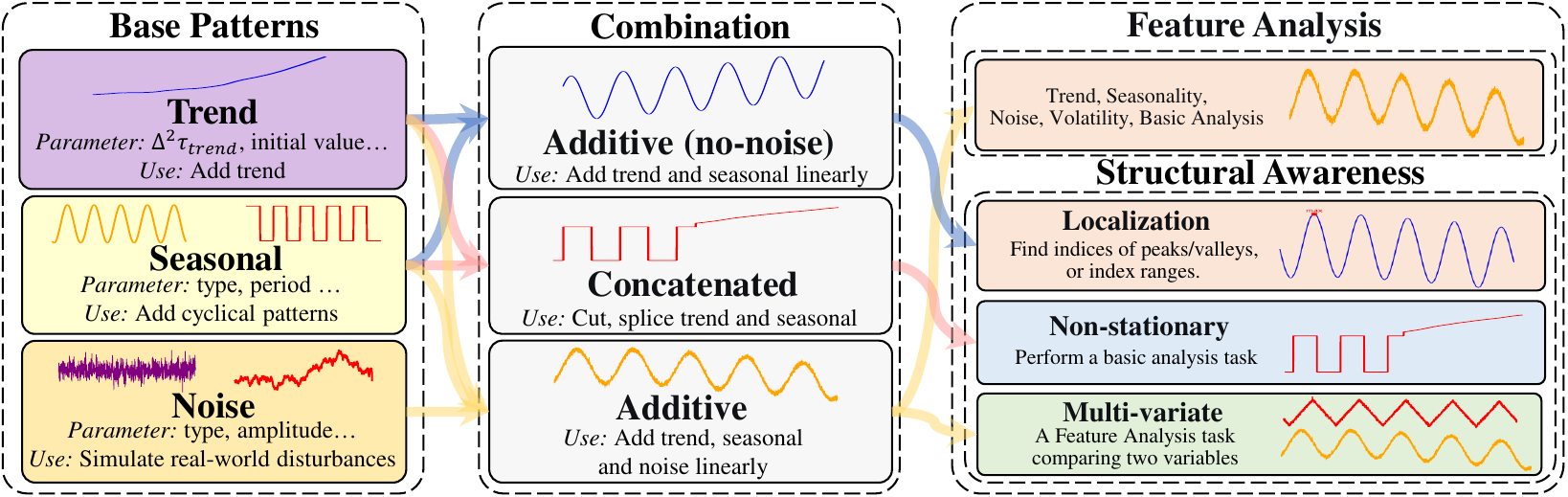}
\caption{\textbf{Base} Construction Pipeline. Synthetic time series with controllable characteristics are generated by concatenating and adding basic components of trend, seasonality, and noise. The plotting style of this figure is adapted from \citep{cai2024timeseriesexam}.}
\label{fig:MMTS-Base_Construction}
\end{figure*}

Although recent work has combined LLMs with time series and released datasets, most efforts remain confined to forecasting \citep{hu2025contextalignment, liu2024timecma, liu2024timemmd, wang2024newsforecast}, while publicly available TSQA datasets are scarce. Moreover, existing TSQA datasets \citep{wang2025chattime, kong2025time, wang2025itformer, xie2024chatts, cai2024timeseriesexam} suffer from domain inconsistency, flat ability taxonomies, and rigid question formats, which isolate different works and hinder cross-comparisons. 


On the univariate side, ChatTime-TSQA \citep{wang2025chattime} relies on fixed templates and targets four basic properties (trend, volatility, seasonality, and outliers). Time-MQA \citep{kong2025time} and Chat-TS \citep{quinlan2025chat} generate QA pairs from real-world data via single-turn prompting, but still provide limited coverage despite manual filtering. For multivariate settings, EngineMT-QA \citep{wang2025itformer} derives QA pairs from aviation-engine time series by combining data analysis with LLM-based refinement and expert validation, yet its narrow domain limits its generality as a benchmark. ChatTS \citep{xie2024chatts} proposes TSEvol-Instruct with iterative prompting over diverse time series, but its flat taxonomy and rigid question formats hinder fine-grained capability evaluation.

In contrast, our proposed \textsc{MMTS-Bench} is a multimodal, multi-dimensional benchmark for TSQA that covers varied difficulty, domains, and both synthetic and real data across univariate and multivariate cases. Through iterative expert curation and human validation, the benchmark offers a balanced and reliable basis for assessing the performance of models.

\section{MMTS-BENCH}

\subsection{Multi-dimensional Task Classification Framework}

To systematically evaluate models’ understanding and reasoning capabilities in time series analysis, we propose a multi-dimensional task classification framework with a corresponding dataset construction methodology. Existing TSQA datasets often lack consistency and hierarchical structure. To address the limitations of such flat classification schemes, we decompose models' temporal understanding into five core dimensions (see Appendix~\ref{sec:appendix DATASETS CLASSIFICATION} for details): structural awareness, feature analysis, temporal reasoning, sequence matching, and cross-modal understanding. For each dimension, we define fundamental tasks: structural awareness involves stationarity and non-stationarity identification, local and global pattern recognition, and univariate and multivariate analysis; feature analysis covers trend, seasonality, volatility, noise, and basic statistics; temporal reasoning includes deductive, inductive, causal, analogical, and counterfactual reasoning; sequence matching focuses on isomorphic, robust, localization, and reverse correspondence; and cross-modal understanding addresses bidirectional mapping between time series and semantic representations. Through combinatorial composition of these tasks, the framework yields 286\footnote{By combining feature analysis and temporal reasoning, 35 composite subtasks are formed. With the addition of structural awareness, this extends to 280. 
Including 4 from sequence matching and 2 from cross-modal understanding, the total reaches 286.} fine-grained composite subtasks.

\begin{table}[h]
\centering
\caption{Overview of five core dimensions, tasks, and related subsets in MMTS-BENCH.}
\label{tab:dimensions_overview}
\resizebox{\columnwidth}{!}{%
\begin{tabular}{l|l|c}
\toprule
\textbf{Dimensions} & \textbf{Task} & \textbf{Subsets} \\
\midrule
Structural Awareness & 
Non-Stationarity, Local-Global, Uni.-Multi. & Base \\
Feature Analysis & 
Trend, Season., Noise, Volat., Basic & Base \& InWild \\
Temporal Reasoning & 
Ded., Ind., Causal, Analog., Count. & InWild \\
Sequence Matching & 
Isomorphic, Robust, Localization, Reverse & Match \\
Cross-Modal Understanding & 
TS$\rightarrow$Semantic, Semantic$\rightarrow$TS & Align \\
\bottomrule
\end{tabular}
}
\end{table}

Based on this framework, we design four subsets under \textsc{MMTS-Bench}. \textbf{Base} provides a controlled synthetic environment focusing on fundamental abilities such as structural awareness and feature analysis, without involving complex reasoning. It includes multiple-choice, binary-choice, and numerical questions. For evaluation, we divide the subset into two splits: the Choice split and the Numerical split. \textbf{InWild} leverages real-world time series to further examine LLMs’ capacity for feature analysis and temporal reasoning under complex and noisy conditions. It consists of multiple-choice and binary-choice questions. \textbf{Match} and \textbf{Align} focus on two less-studied dimensions in time series analysis, namely similarity matching and cross-modal understanding, and are composed of multiple-choice questions. Table \ref{tab:dimensions_overview} summarizes the subtasks across the five core dimensions and their links to the subsets, with full definitions given in Table \ref{tab:full_definition} (Appendix~\ref{sec:appendix DATASETS CLASSIFICATION}). This multi-dimensional construction enables fine-grained profiling of LLM capabilities in time series analysis and helps identify the bottlenecks they face in complex analytical reasoning tasks.

\subsection{Synthetic Dataset}

\begin{figure*}[h]
\centering
\includegraphics[width=0.95\textwidth]{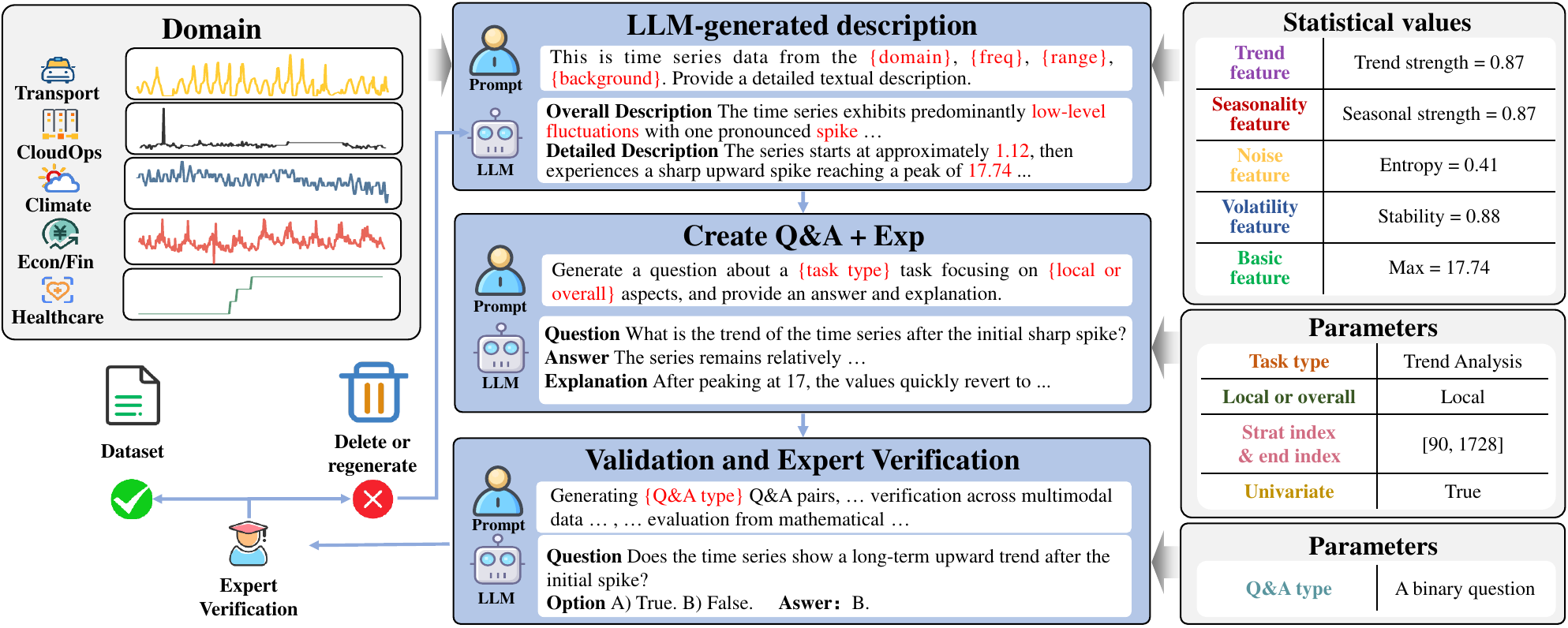}
\caption{\textbf{InWild} Construction Pipeline. Flowchart illustrating the conversion of domain-specific time series via statistical analysis and multimodal input preparation for LLMs. It highlights the feedback loop between LLM generation and expert verification, showing how raw sequences from multiple domains (e.g., Economics, Transport, CloudOps, …) are enriched with features (e.g., trend strength, seasonal strength, entropy, …) and structured as inputs for automated QA generation.}
\label{fig:MMTS-InWild Construction}
\end{figure*}


\textbf{Base} is designed to enable fundamental evaluations under controllable conditions using synthetic time series, while being clearly differentiated from prior modular benchmarks. It employs 17 expert-designed templates with explicit, continuous parameters to generate QA pairs over interpretable sequences, consistent with established synthesis practice \citep{cai2024timeseriesexam, das2024decoderonlyfoundationmodeltimeseries, fu2024synthetictimeseriesdatareally, tadnet}. Unlike earlier work that follows a relatively fixed compositional pipeline and largely relies on qualitative annotations, our framework adopts a flexible STL-inspired decomposition with multiple composition modes (including additive and concatenated synthesis), uses richer trend/noise processes beyond simple functional forms, and records the exact numerical parameters used in construction. This parameter-level logging enables graded questions and fine-grained sensitivity tests over time series components (e.g., distinguishing weak vs. strong seasonality), while the concatenated mode explicitly models non-stationarity via controlled regime shifts. These design choices jointly support richer and more diagnostic task construction in \textbf{Base} (detailed mathematical formulation in Appendix~\ref{sec:appendix SYNTHETIC DATASET}).

\subsection{Real-World Datasets}


\textbf{InWild} targets realistic and complex time series scenarios to assess models’ real-world generalization as well as their analytical and reasoning capabilities, complementing \textbf{Base}, which emphasizes controllability and fundamental property evaluation. We construct \textbf{InWild} using a principled, three-stage, reasoning-oriented pipeline with LLM (Figure \ref{fig:MMTS-InWild Construction}; Appendix~\ref{sec: appendix REAL-WORLD DATASET}). \textbf{(1) Multimodal Context Integration and Description Generation.} Given raw sequences, visualizations, domain metadata that grounds physical semantics (e.g., hourly traffic volume, CPU/memory utilization), and pre-computed statistical features, the LLM produces coherent global summaries and fine-grained local descriptions. \textbf{(2) Reasoning-Driven QA Synthesis.} Conditioned on predefined initialization parameters (task type, global vs. local scope, uni- vs. multivariate setting, and the target index range), the model converts the contextualized descriptions into Question–Answer–Explanation (Q-A-E) triples, explicitly requiring explanations to justify conclusions rather than relying on templated, single-turn generation. \textbf{(3) Multi-Dimensional Verification and Expert Refinement.} Each instance is checked for logical validity, mathematical correctness, and cross-modal consistency between numerical and visual representations; the resulting QA pairs are then standardized and further reviewed by human experts to ensure reliability and domain fidelity.

\begin{figure*}[h]
\centering
\includegraphics[width=0.9\textwidth]{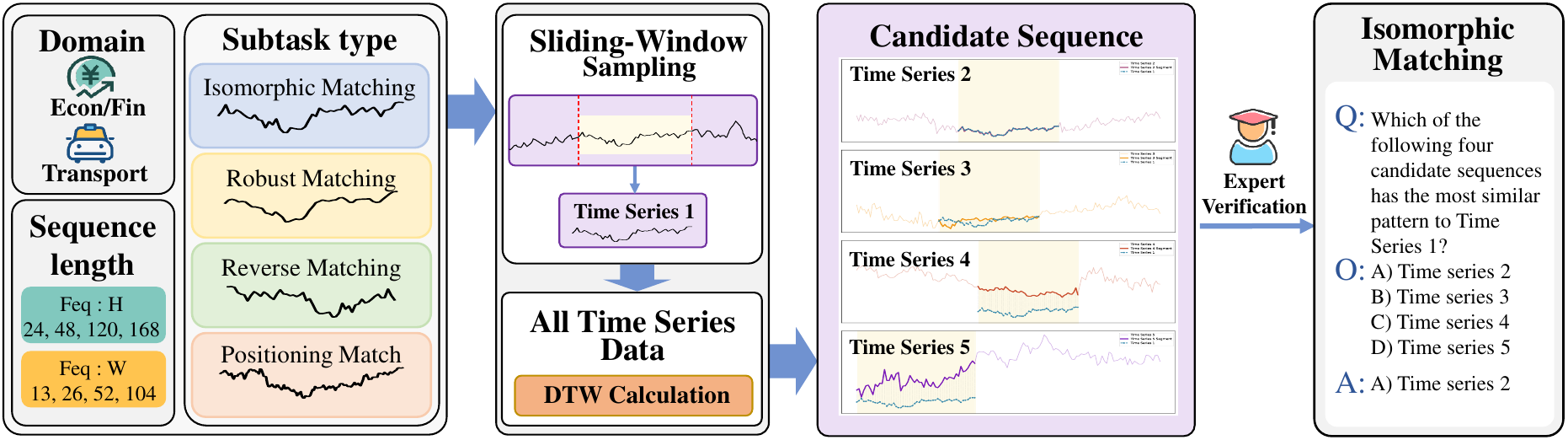}
\caption{\textbf{Match} Construction Pipeline. Fragments are extracted from real-world time series, and candidate series with different similarity levels are retrieved using DTW. QA pairs are then formed with fixed templates, while transformations such as smoothing, extension, and reversal create four task paradigms of varying difficulty.}
\label{fig:MMTS-Match_Construction}
\end{figure*}

\textbf{Match} is designed to evaluate models’ ability to perform similarity matching on time series. It is constructed by extracting fragments from real-world time series and applying Dynamic Time Warping (DTW) to search for four candidate sequences at different similarity levels. Each fragment and its candidate sequences are then combined into a multiple-choice question, where a template asks the model to identify the candidate most similar to the fragment (Figure \ref{fig:MMTS-Match_Construction}). In addition, we apply operations such as smoothing, extension, and reversal to the fragment samples, constructing four task paradigms of different difficulty levels (details in Appendix~\ref{sec: appendix REAL-WORLD DATASET}): (1) Isomorphic, matching sequences with identical lengths, (2) Robust, matching sequences after smoothing, (3) Localization, localizing target patterns within longer sequences, and (4) Reverse, matching sequences after reversal. 

\textbf{Align} evaluates models’ cross-modal alignment between time series and natural-language descriptions. We reuse descriptions generated in the first stage of \textbf{InWild} and pair each with its corresponding time series to form multiple-choice instances (details in Appendix~\ref{sec:appendix ALIGN SUBSET}). This yields two symmetric tasks: (1) \emph{Time series $\rightarrow$ semantic (TS$\to$Sem)}, where the model selects the correct description for a given series from a set of candidates; and (2) \emph{Semantic $\rightarrow$ time series (Sem$\to$TS)}, where the model identifies the matching series for a given description among candidate sequences. To increase difficulty, distractor series are drawn from three real-world domains with overlapping value ranges, making incorrect options plausible.

\subsection{Human-In-The-Loop Curation}

To ensure high dataset quality, we adopt systematic human-in-the-loop curation \citep{wu2022survey} across all subsets, carried out by a fixed team of ten time series domain experts. The review criteria are subset-specific: for \textbf{InWild}, experts assess question soundness and reasoning validity; for \textbf{Match}, experts visually confirm that the DTW-nearest candidate is also the most similar in shape, aligning metric-based retrieval with human perception; and for \textbf{Align}, experts verify that each description faithfully reflects the underlying temporal patterns. We further perform statistical validation to assess annotation reliability: Fleiss’ $\kappa$ \citep{fleiss1971measuring} reaches 0.73, indicating substantial inter-annotator agreement \citep{landis1977measurement}. Final ground-truth labels are determined by majority vote, with ambiguous cases escalated to a second-round adjudication.

To motivate the design of our \textbf{InWild} three-stage pipeline, we compare it against one-step prompting and examine how performance changes as we progressively add structured constraints and human curation. Without time series inputs, one-step prompting yields about 57\% accuracy (vs. 37.5\% expected from random guessing given the mix of multiple-choice and binary questions). Using our three-stage pipeline without expert review reduces this to 44\%, and adding expert review further lowers it to 35\%. This trend suggests that the staged construction better constrains generation and, together with expert filtering, reduces spurious cues and potential answer leakage that can otherwise inflate apparent accuracy.

\subsection{Statistical Reliability and Validity Analysis}
\label{sec:statistical_analysis}

To complement human curation, we conduct statistical analyses to verify the benchmark’s reliability and validity (see Appendix~\ref{sec: robustness}). Specifically, we run two experiments: (1) \textbf{Evaluation Stability:} using bootstrap confidence intervals and iterative subsampling, \textsc{MMTS-Bench} yields consistently narrow intervals with a coefficient of variation on the order of $10^{-3}$, indicating strong robustness to sampling variance; and (2) \textbf{Validity against Shortcut Learning:} to ensure performance reflects intrinsic time series reasoning rather than superficial artifacts, we analyze accuracy against surface attributes such as sequence length and dimensionality, observing negligible correlations (e.g., $|r|<0.08$ for sequence length) and minimal performance gaps, suggesting \textsc{MMTS-Bench} is largely insensitive to these factors and resists shortcut learning.

\section{Evaluation Results}
Using the \textsc{MMTS-Bench} dataset, we conducted a systematic benchmarking and analysis of the latest open-source and closed-source LLMs alongside state-of-the-art (SOTA) TS-LLMs. We report \emph{Accuracy}, \emph{Accuracy@N\%}, and \emph{Relative Accuracy} (definitions in Appendix~\ref{sec:metrics}) across multiple subsets spanning different task dimensions. To ensure statistical robustness and experimental reliability, all experiments were conducted five times independently at temperature $1.0$, with results averaged across trials.

\textbf{Model Selection for Evaluation.} We conduct a comprehensive evaluation on \textsc{MMTS-Bench} using three representative categories of LLMs to assess their performance on TSQA tasks. Our selection includes: (1) Closed-source models: Claude 3.7 Sonnet \citep{anthropic2025claude37sonnet}, Claude Sonnet 4 \citep{anthropic2025claude4}, Gemini 2.5 Flash/Pro \citep{comanici2025gemini}, GPT-5 Minimal/High \citep{openai2025gpt5}, GPT-4.1/4.1 mini \citep{openai2025gpt41}, and GPT-4o \citep{openai2024hello_gpt4o}; (2) Open-source models: DeepSeek V3 \citep{deepseekai2024deepseekv3technicalreport}, Kimi K2 \citep{kimiteam2025kimik2openagentic}, and Qwen series \citep{qwen2.5,qwen3} (including 2.5 and 3 variants with different parameter scales). Notably, for the Qwen3 series, we evaluate both thinking and non-thinking modes. All evaluated Qwen2.5 models are instruction-tuned; we omit this designation in subsequent mentions for brevity; (3) TS-LLMs: ChatTS \citep{xie2024chatts}, ITFormer \citep{wang2025itformer}, and ChatTime \citep{wang2025chattime}, which are specifically designed for time series data analysis. For MLLMs (Qwen2.5-VL \citep{Qwen2.5-VL}, Claude Sonnet series \citep{anthropic2025claude37sonnet, anthropic2025claude4}, Gemini 2.5 Pro \citep{comanici2025gemini}, GPT-4 series \citep{openai2023gpt4_research}), we evaluate across text-only, vision-only, and vision-text combined inputs. To ensure a fair evaluation, we design a standardized time series input format (see Appendix \ref{sec:Standardized TS Input Format} for details); for TS-LLMs, we carefully reuse the original inference scripts. This multi-dimensional evaluation framework aims to clarify the impact of different model architectures and input modalities on time series understanding and reasoning tasks.

\begin{table}[h]
\centering
\caption{Performance of representative LLMs on MMTS-BENCH. TS denotes time series. Best results are \textbf{\underline{bold and underlined}}.}
\label{tab:model_MMTS_performance}
\resizebox{\columnwidth}{!}{%
\renewcommand{\arraystretch}{0.9}
\setlength{\tabcolsep}{3pt}
\begin{tabular}{l|lc|c|cccc}
\toprule
\multirow{2}{*}{\textbf{Cat.}} & \multirow{2}{*}{\textbf{Model}} & \multirow{2}{*}{\textbf{Type}} & \multirow{2}{*}{\textbf{Avg.}}
& \multicolumn{4}{c}{\textbf{MMTS-BENCH Subsets}} \\
\cmidrule(lr){5-8}
& & & & \textbf{Base} & \textbf{InWild} & \textbf{Match} & \textbf{Align} \\
\midrule
\multirow{4}{*}{Closed}
& GPT-5 High & Text & \textbf{\underline{0.74}} & \textbf{\underline{0.51}} & \textbf{\underline{0.72}} & \textbf{\underline{0.82}} & \textbf{\underline{0.99}} \\
& GPT-4o & Text & 0.61 & 0.42 & 0.62 & 0.50 & 0.97 \\
& Claude Sonnet 4 & Text & 0.71 & 0.49 & 0.71 & 0.71 & 0.98 \\
& Gemini 2.5 Pro & Text & 0.70 & 0.48 & 0.68 & 0.79 & 0.98 \\
\midrule
\multirow{4}{*}{Open}
& Kimi K2 & Text & \textbf{\underline{0.63}} & \textbf{\underline{0.45}} & \textbf{\underline{0.63}} & 0.60 & \textbf{\underline{0.95}} \\
& DeepSeek V3 & Text & 0.62 & 0.41 & 0.61 & \textbf{\underline{0.65}} & \textbf{\underline{0.95}} \\
& Qwen2.5-14B & Text & 0.55 & 0.35 & 0.53 & 0.57 & 0.88 \\
& Qwen2.5-7B & Text & 0.45 & 0.33 & 0.44 & 0.40 & 0.69 \\
\midrule
\multirow{2}{*}{TS}
& ChatTS & TS & \textbf{\underline{0.49}} & \textbf{\underline{0.39}} & \textbf{\underline{0.50}} & \textbf{\underline{0.37}} & \textbf{\underline{0.80}} \\
& ITFormer & TS & 0.31 & 0.31 & 0.33 & 0.24 & 0.29 \\
\bottomrule
\end{tabular}%
}
\end{table}

\subsection{Vertical Comparison of LLMs}

\textbf{TS-LLMs Show Limited Generalization Capabilities.} From Table \ref{tab:model_MMTS_performance} and the more comprehensive experimental results in the Appendix \ref{sec:appendix:Full Results}, we observe that both closed-source and open-source general-purpose LLMs consistently outperform TS-LLMs across diverse subtasks in \textsc{MMTS-Bench}, including \textbf{InWild}, \textbf{Match}, \textbf{Align}, and the Choice split in \textbf{Base}. TS-LLMs show marked weaknesses in OOD generalization: ChatTS is comparable to, or slightly below, its base model Qwen2.5-14B; ITFormer lags substantially behind Qwen2.5-7B when applied outside the aero-engine domain; and ChatTime fails to produce valid outputs. These findings indicate that while TS-LLMs may perform adequately within narrow domains, their generalization remains severely limited. 

\textbf{Existing Multimodal Alignment in TS-LLMs Remains Inefficient.}
 To investigate the key factors affecting TS-LLM performance, we modified the training pipeline of the current SOTA model ChatTS and conducted a series of ablation studies on encoder architecture, scale, positional encoding, backbone LLM size, and prompt prefix. The results show that model performance is predominantly determined by the backbone LLM size, while being largely insensitive to encoder structure, scale, and positional encoding, suggesting that the encoder’s contribution remains limited and underdeveloped. Notably, augmenting the prompt with simple statistical summaries leads to substantial improvements in reasoning accuracy, underscoring the importance of task-aware prompt design. Details are provided in Section~\ref{ablation_study}.

\subsection{Cross-Dimensional Comparison of LLMs Performance}

\begin{figure}[htbp]
\begin{center}
\includegraphics[width=\columnwidth]{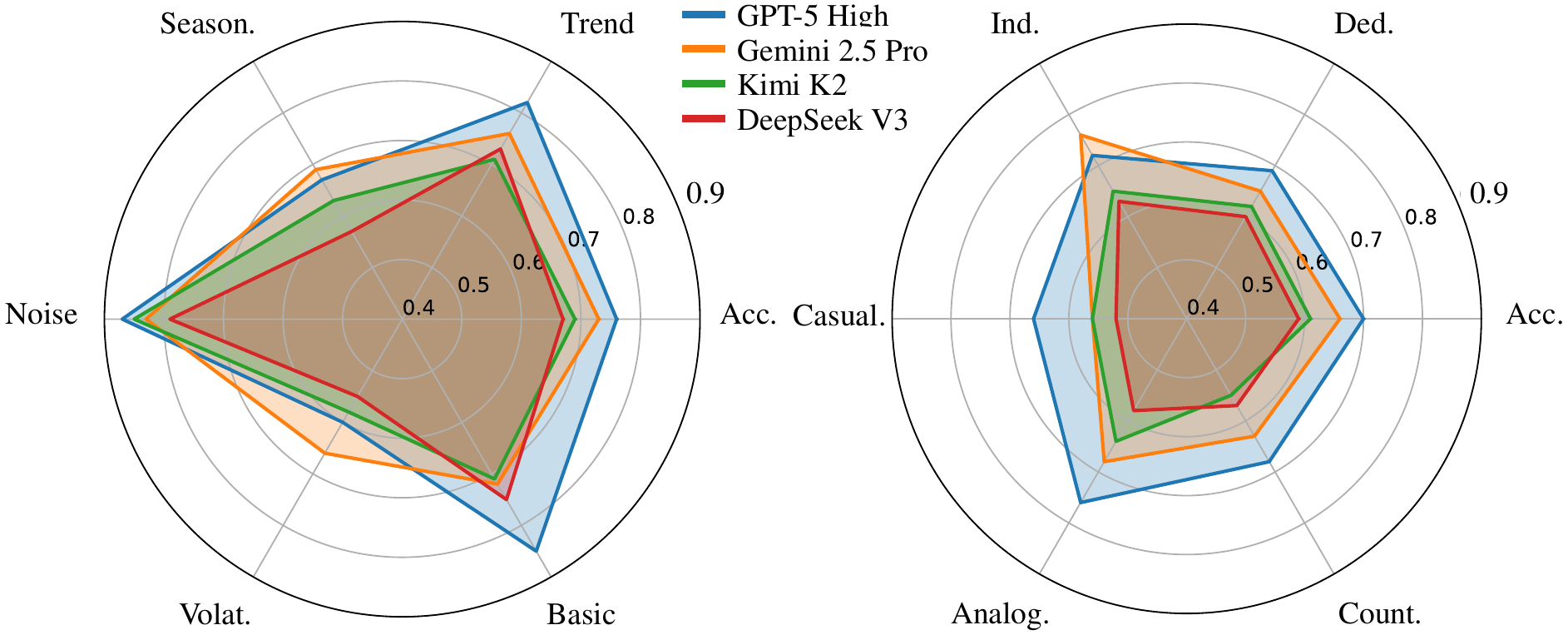}
\end{center}
\caption{Task-level accuracy of top-ranked LLMs on \textbf{InWild} (left: feature analysis; right: temporal reasoning).}
\label{fig:InWild_radar_text}
\end{figure}

\textbf{LLMs Underperform on Temporal Reasoning Relative to Feature Analysis.} As shown in Table \ref{tab:MMTS-InWild-performance} (Appendix~\ref{sec:appendix:Full Results}), current LLMs reach an average accuracy of 62\% on feature analysis tasks in \textbf{InWild}, notably higher than the 55\% achieved on temporal reasoning tasks. A comparison under unified input modalities indicates that Claude Sonnet 4 and Gemini 2.5 Pro are the strongest closed-source LLMs\footnote{Claude 3.7 Sonnet ranked second but was excluded due to its substantial involvement in dataset construction.}, while DeepSeek V3 and Kimi K2 lead among open-source LLMs (Figure \ref{fig:InWild_radar_text}). The results further indicate that weaker performance on feature analysis generally coincides with poor temporal reasoning. For instance, GPT-4.1 Mini among closed-source models and Qwen2.5-7B among open-source models both follow this trend, suggesting that insufficient feature analysis capability constrains temporal understanding.

Within the Feature Analysis dimension, seasonality tasks consistently yield the lowest accuracy across \textbf{Base} and \textbf{InWild}. This indicates that LLMs may struggle with capturing seasonal patterns, making seasonality a particularly challenging task. In the Temporal Reasoning dimension, results further reveal that causal and counterfactual reasoning are especially difficult compared to other reasoning tasks. As illustrated in Figure \ref{fig:error_case} (Appendix~\ref{sec: appendix REAL-WORLD DATASET}), LLMs frequently fall into local reasoning traps in causal tasks, failing to capture causal relations from a global perspective. Similarly, counterfactual reasoning involves reconstructing dependencies under hypothetical conditions. When LLMs lack sufficient local and global awareness, they tend to make faulty conditional assumptions, which ultimately leads to reasoning failures.

\textbf{LLMs Show Weaknesses in Local Tasks Compared to Global Tasks.} As shown in the \textbf{Base} results (Table \ref{tab:MMTS-Base-consolidated}), LLMs show their most pronounced weaknesses in Local subtasks under the structural awareness dimension. Across all model categories, accuracy on local tasks remains low, whereas performance on global tasks is considerably higher. This suggests that while LLMs can capture overall regularities in time series structures, they generally lack precise localization and fine-grained discrimination across different input modalities—an issue that connects with the reasoning failures discussed above. In contrast, performance gaps in non-stationarity and univariate–multivariate subtasks remain relatively modest. Non-stationary sequences are typically constructed by concatenating subsequences with distinct statistical properties, leading to global shifts that can be detected through overall structural and distributional cues. 


\begin{table}[htbp]
\centering
\caption{The \textit{Accuracy@10\%} metric of different models on the \textbf{Base} subset's numerical split.}
\label{tab:MMTS-Base-consolidated}
\resizebox{\columnwidth}{!}{%
\renewcommand{\arraystretch}{0.9}
\setlength{\tabcolsep}{3pt}
\begin{tabular}{l|lc|cc|cc|cc}
\toprule
\multirow{2}{*}{\textbf{Cat.}} & \multirow{2}{*}{\textbf{Model}} & \multirow{2}{*}{\textbf{Type}} 
& \multicolumn{6}{c}{\textbf{Structural Awareness}}\\
\cmidrule(lr){4-9}
& & & Stat. & Non-S. & Local & Over. & Uni. & Multi. \\
\midrule
\multirow{3}{*}{Open}
& Kimi-K2 & Text & 0.56 & 0.56 & 0.19 & 0.45 & 0.56 & 0.54\\
& Qwen3-32b$^{cot}$ & Text & 0.40 & 0.40 & 0.20 & 0.32 & 0.40 & 0.33 \\
& Qwen3-32b & Text & 0.47 & 0.50 & 0.07 & 0.38 & 0.47 & 0.42 \\
\midrule
\multirow{3}{*}{Closed}
& GPT-4o & Text & 0.53 & \textbf{\underline{0.58}} & 0.13 & 0.43 & 0.53 & 0.51 \\
& GPT-4o & V & 0.34 & 0.29 & \textbf{\underline{0.31}} & 0.31 & 0.34 & 0.34 \\
& GPT-4o & V+T & \textbf{\underline{0.59}} & 0.57 & \textbf{\underline{0.31}} & \textbf{\underline{0.56}} & \textbf{\underline{0.59}} & \textbf{\underline{0.56}} \\
\midrule
TS & ChatTS & TS & 0.41 & 0.42 & 0.12 & 0.40 & 0.41 & 0.40 \\
\bottomrule
\end{tabular}
}
\end{table}

Moreover, in the evaluations on the \textbf{Match} and \textbf{Align} subsets (see Table \ref{tab:MMTS-Match-performance} and Table \ref{tab:MMTS-Align-performance-2} in Appendix~\ref{sec:appendix:Full Results}), we observe two key findings: (1) among the four sequence matching tasks, LLMs perform significantly worse on Localization Matching and Reverse Matching than on the remaining ones. This may be due to the inherent limitations of attention mechanisms and autoregressive paradigms under temporal direction transformations, though further experiments are required to confirm this; (2) in the two tasks under the Cross-Modal dimension, LLMs generally exhibit strong performance. This is because LLMs, by leveraging basic statistical cues (e.g., maxima, minima) and conducting logical reasoning, can align sequences with textual descriptions.

\subsection{Analysis of Approaches for Enhancing LLM Performance on Time Series Tasks}
\begin{figure}[t]
    \centering
    \begin{subfigure}{\columnwidth}
        \centering
        \includegraphics[width=0.85\columnwidth]{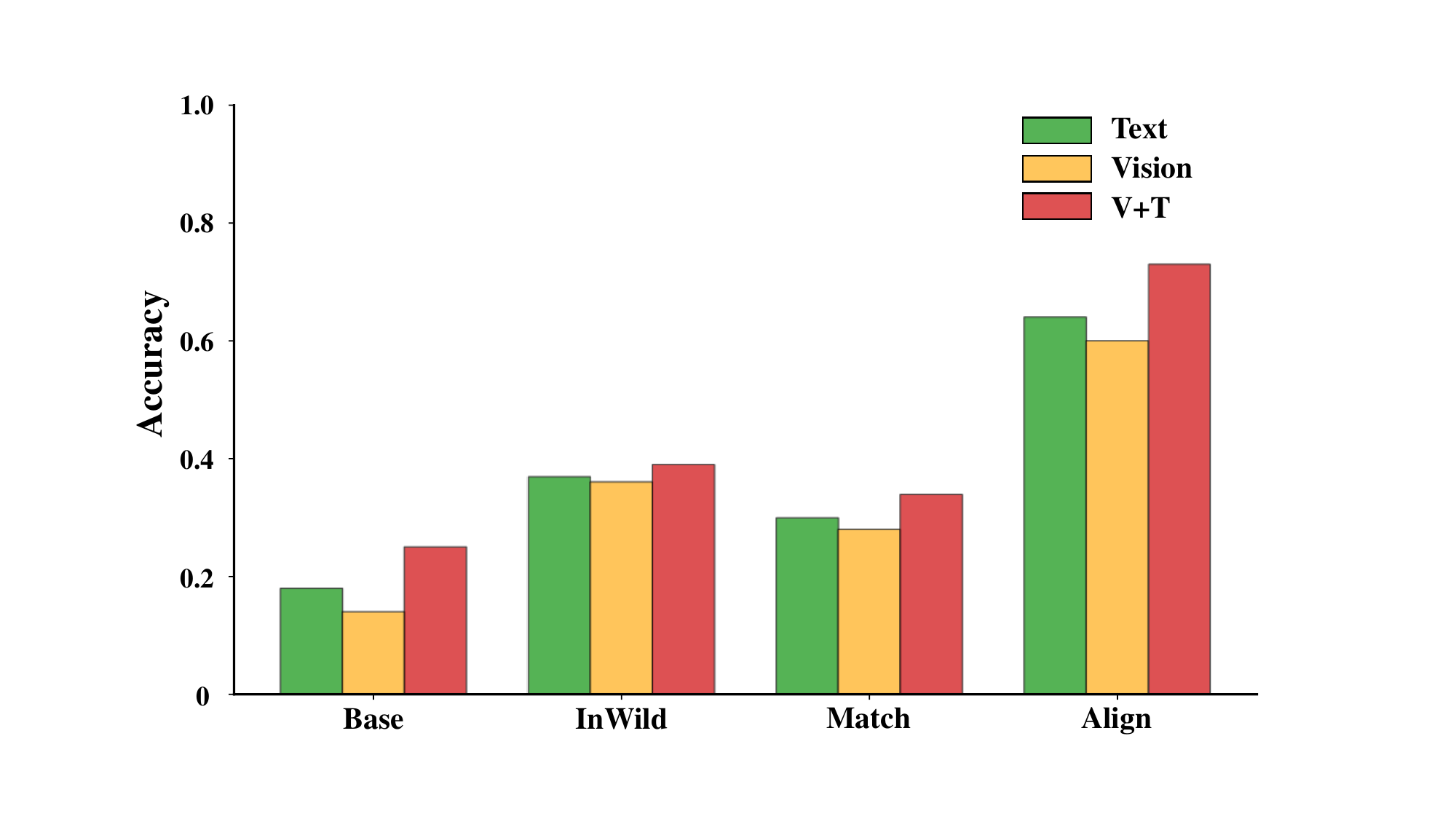}
        \caption{Performance of Qwen2.5-VL-7B across four subsets}
        \label{fig:qwen2.5 on MMTS}
    \end{subfigure}
    
    
    \begin{subfigure}{\columnwidth}
        \centering
        \includegraphics[width=0.85\columnwidth]{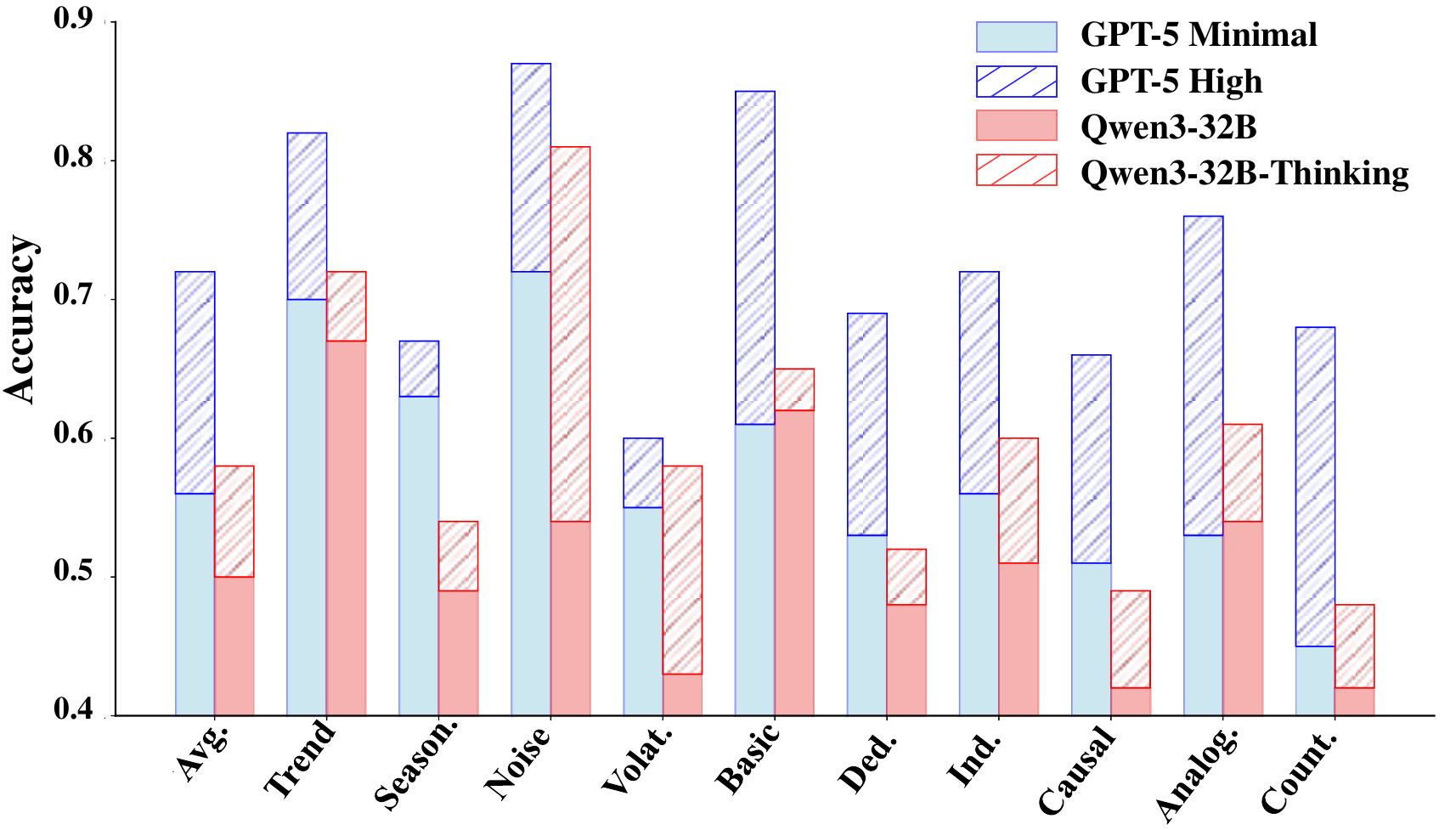}
        \caption{Accuracy comparison under CoT and non-CoT settings}
        \label{fig:thinking_none}
    \end{subfigure}
    
    \caption{(a) Evaluation results of Qwen2.5-VL-7B across four subsets. (b) Accuracy gains of GPT-5 and Qwen3-32B on \textbf{InWild} across different subtasks with and without CoT.}
    \label{fig:Multi-bar chart}
\end{figure}

\textbf{Multimodal Fusion Enhances LLMs’ Ability in Time Series Analysis.} As shown in Table \ref{tab:MMTS-InWild-performance} (Appendix~\ref{sec:appendix:Full Results}), Gemini 2.5 Pro achieves 68\% accuracy with text-only inputs, 72\% with vision-only inputs, and reaches 76\% when combining modalities. Similarly, GPT-4.1 exhibits similar improvements with multimodal fusion. Building on this observation, we further evaluated Qwen2.5-VL-7B across \textsc{MMTS-Bench}, which also showed consistent gains (Figure \ref{fig:qwen2.5 on MMTS}). However, such improvements are not universal; for example, GPT-4.1 Mini shows reduced performance on \textbf{InWild}. We suspect this limitation arises from differences in the LLMs’ inherent ability to integrate multiple modalities. Therefore, introducing additional modalities is an effective way to narrow the information gap of LLMs in time series analysis, but its effectiveness mainly depends on the fusion design and training methods of the LLMs.

\textbf{CoT Reasoning Enhances LLMs Beyond Parameter Scaling in Time Series Analysis.} Experiments with the Qwen2.5 series in Table \ref{tab:MMTS-InWild-performance} (Appendix~\ref{sec:appendix:Full Results}) demonstrate a scaling-law effect \citep{kaplan2020scalinglawsneurallanguage}: scaling from 7B to 14B parameters yields clear performance gains, but further growth to 32B provides marginal improvements, indicating diminishing returns for temporal reasoning tasks in time series analysis. In contrast, enabling CoT reasoning produces substantial improvements across all \textbf{InWild} subtasks for Qwen3 and GPT-5, as shown in Figure \ref{fig:thinking_none}. The benefits are especially pronounced in temporal reasoning, where the average improvement surpasses that observed in feature analysis tasks, and additional evaluations on the other subsets confirm this trend. These findings highlight that scaling laws impose inherent limits on parameter-based gains, whereas activating CoT reasoning enables models to capture temporal dependencies more effectively. Therefore, future work should focus on leveraging CoT reasoning to enhance LLMs’ performance on time series analysis, rather than relying solely on scaling up LLM parameter size.

\section{Ablation Study}
\label{ablation_study}
Across \textsc{MMTS-Bench}, ChatTS \cite{xie2024chatts} is the strongest open-source TS-LLM baseline, demonstrating robust time series analysis and reasoning ability. To understand what drives this performance and to lead future TS-LLM design, we reproduce and modify the official ChatTS training pipeline using the released training data and recommended training strategy, conducting controlled ablations over the time series encoder (architecture, size, and positional encoding), the LLM backbone scale, and the statistical prompt prefix. Experimental settings are provided in Appendix~\ref{appen:ablation_study}.

\subsection{Encoder Architecture Has Limited Impact}
\label{sec:ablation_encoder_arch}

We vary the encoder architecture while keeping the overall training setup fixed, comparing an MLP (17.1M), a CNN (50.4M), and a Transformer (6.3M). As shown in Table~\ref{tab:ablation_encoder}, performance differences across architectures are marginal, suggesting that \emph{encoder type is not the main performance bottleneck} under the current alignment paradigm. Relative to the Qwen2.5-3B baseline (treating time series as plain text), adding an encoder yields little change on \textbf{InWild} and \textbf{Match}, but improves Sem$\to$TS while degrading TS$\to$Sem, indicating a potential directional bias induced by the encoder–training data interaction.


\begin{table}[htbp]
\centering
\caption{Performance of TS-LLMs with different time series encoder architectures. The baseline is Qwen2.5-3B-Instruct.}
\label{tab:ablation_encoder}
\resizebox{\columnwidth}{!}{%
\begin{tabular}{l|c|ccc}
\toprule
\textbf{Dataset} & \textbf{Baseline} & \textbf{MLP(17.1M)} & \textbf{CNN(50.4M)} & \textbf{Trans.(6.3M)} \\
\midrule
InWild & 0.39 & 0.38 & 0.37 & 0.39 \\
Match & 0.27 & 0.31 & 0.28 & 0.30 \\
Sem$\to$TS & 0.49 & 0.59 & 0.61 & 0.61 \\
TS$\to$Sem & 0.64 & 0.46 & 0.44 & 0.48 \\
\bottomrule
\end{tabular}%
}
\end{table}

\subsection{Backbone Scale Dominates, Encoder Doesn't}
\label{sec:ablation_encoder_scale}

To further examine scaling effects within a fixed architecture, we evaluate MLP encoders with varying depths (1, 3, 5, and 7 layers), all built on the same Qwen2.5-3B-Instruct backbone, as reported in Table \ref{tab:ablation_layers}. We additionally compare these models with larger-backbone baselines, including Qwen2.5-14B-Instruct and the ChatTS model. Notably, ChatTS adopts Qwen2.5-14B-Instruct as its backbone together with a 5-layer MLP as the time series encoder. Results show that increasing the number of MLP layers does not yield a monotonic improvement, indicating that \emph{simply deepening the encoder does not reliably improve performance}.
Results on aligned backbones further mirror the earlier directional pattern: Sem$\to$TS improves substantially, whereas TS$\to$Sem degrades relative to the instruction-tuned baseline, reinforcing the observation that the current training setup may encourage asymmetric alignment behavior.

\begin{table}[t]
\centering
\caption{Ablation on time series encoder depth.}
\label{tab:ablation_layers}
\resizebox{\columnwidth}{!}{%
\setlength{\tabcolsep}{2.5pt}
\begin{tabular}{l|c|cccc|cc}
\toprule
\textbf{Data} & \textbf{3B} & \textbf{1L(0.3M)} & \textbf{3L(8.7M)} & \textbf{5L(17.1M)} & \textbf{7L(25.5M)} & \textbf{14B} & \textbf{ChatTS} \\
\midrule
InWild & 0.39 & 0.39 & 0.38 & 0.38 & 0.37 & 0.53 & 0.50 \\
Match & 0.27 & 0.31 & 0.30 & 0.31 & 0.33 & 0.57 & 0.37 \\
Sem$\to$TS & 0.49 & 0.61 & 0.52 & 0.59 & 0.61 & 0.87 & 0.91 \\
TS$\to$Sem & 0.64 & 0.45 & 0.44 & 0.46 & 0.45 & 0.89 & 0.68 \\
\bottomrule
\end{tabular}%
}
\end{table}

\subsection{Positional Encoding Has Only Minor Effects}
\label{sec:ablation_positional}

We then evaluate positional encoding strategies used in the official code, including no positional encoding, learnable position embeddings, and normalized index values. Table~\ref{tab:ablation_pos_embs} shows only small variations across configurations, indicating that \emph{positional encoding choice contributes limited gains} compared with backbone scaling or other factors.

\begin{table}[h!]
\centering
\caption{Performance of models with different positional encoding strategies. \texttt{no\_emb} denotes no positional encoding, \texttt{pos\_emb} denotes learnable embeddings, and \texttt{pos\_idx} denotes normalized index values used as positional encoding.}
\label{tab:ablation_pos_embs}
\setlength{\tabcolsep}{8pt}
\begin{tabular}{l | c c c}
\toprule
\textbf{Dataset} & \textbf{no\_emb} & \textbf{pos\_emb} & \textbf{pos\_idx} \\
\midrule
InWild & 0.39 & 0.38 & 0.39 \\
Match & 0.31 & 0.30 & 0.29 \\
Sem$\to$TS & 0.59 & 0.52 & 0.52 \\
TS$\to$Sem & 0.42 & 0.44 & 0.45 \\
\bottomrule
\end{tabular}
\end{table}

\subsection{Statistical Prompt Prefix Is a Key Lever}
\label{sec:ablation_prefix}

Finally, we study the statistical prompt prefix in ChatTS, which encodes basic descriptors of the input series (e.g., offset, scale factor, length, min/max, and boundary values). We compare training and evaluating with the prefix (\textbf{ON}) versus without it (\textbf{OFF}), and also test models trained without the prefix but evaluated with it (\textbf{OFF}$^{*}$). Table~\ref{tab:ablation_preifx} shows that the prefix yields substantial gains, especially on the two alignment tasks, and that providing it at inference time largely recovers performance. This suggests that \emph{explicit statistical information serves as an effective auxiliary signal} that improves interpretability and reduces reasoning barriers during inference.

\begin{table}[h!]
\centering
\caption{Performance of models trained with and without the prompt prefix.}
\label{tab:ablation_preifx}
\setlength{\tabcolsep}{12pt}
\begin{tabular}{l | c c c}
\toprule
\textbf{Dataset} & \textbf{ON} & \textbf{OFF} & \textbf{OFF$^{*}$} \\
\midrule
InWild & 0.38 & 0.37 & 0.37 \\
Match & 0.31 & 0.26 & 0.33 \\
Sem$\to$TS & 0.59 & 0.24 & 0.60 \\
TS$\to$Sem & 0.46 & 0.21 & 0.46 \\
\bottomrule
\end{tabular}
\end{table}

\subsection{Summary of Findings}
\label{sec:ablation_summary}

Overall, our ablations lead to three takeaways. (1) \textbf{Encoder effects are limited}: encoder architecture, depth, and positional encoding yield only minor changes under the current training paradigm, indicating that more effective time-series--text alignment mechanisms remain an open challenge. (2) \textbf{Backbone scale dominates}: scaling the LLM backbone consistently improves performance across tasks, highlighting the central role of general reasoning capacity. (3) \textbf{Prompt design matters}: the statistical prefix provides strong, practical gains, suggesting that lightweight auxiliary signals and prompt engineering are promising directions for strengthening TS-LLM reasoning.

\section{Conclusion}
We presented \textsc{MMTS-Bench}, a comprehensive benchmark comprising 2,424 TSQA pairs across four specialized subsets for evaluating multi-modal time series understanding and reasoning abilities. Our extensive evaluations reveal that general-purpose LLMs outperform TS-LLMs in cross-domain generalization, and that current LLMs struggle with fine-grained localization and complex reasoning. We also find that simply scaling model size yields diminishing returns; instead, performance is more effectively enhanced through multi-modal inputs and explicit reasoning strategies like CoT, with the backbone LLM's capability being the dominant success factor. Complementary ablation studies suggest that, under current training paradigms, commonly adopted encoder designs provide limited gains, which in turn highlights the need for deeper investigation into TS-LLM architectures and more effective time-series--text alignment mechanisms beyond straightforward encoder scaling.


\textbf{Limitations and Future Work.} \textsc{MMTS-Bench} currently does not cover several traditional time series evaluation tasks, such as forecasting, imputation, anomaly detection, and representation learning, which are also important components of time series capability. The benchmark also relies heavily on English annotations, potentially missing insights from multilingual temporal reasoning. Future work includes: (1) extending to longer-horizon sequences beyond current context limits, (2) adding complementary suites for traditional time series tasks (forecasting, imputation, anomaly detection, and representation learning) to provide a more complete picture of model capability, (3) developing more effective time series–text alignment paradigms beyond current encoder architectures, and (4) exploring prompt and tool-use strategies that leverage statistical features more systematically. We hope this evaluation protocol will help steer the development of time series LLMs toward robust, generalizable time series understanding.

\bibliographystyle{ACM-Reference-Format}
\bibliography{sample-base}

\appendix

\clearpage

\section*{Table of Contents}

\begin{itemize}
    \item Appendix~\ref{sec:Details About MMTS-Bench}: Details About \textsc{MMTS-Bench}. Definitions of evaluation metrics, sources of real-world time series data, dataset classification, and dataset construction methods. 
    \item Appendix~\ref{sec:Standardized TS Input Format}: Evaluation and Experimental Setup. Standardized time series input format, evaluation prompts, time series visualization, and experimental setup for ablation study.
    \item Appendix~\ref{sec:appendix:Full Results}: Full Results. Complete testing results of LLMs evaluated with \textsc{MMTS-Bench}. 
    \item Appendix~\ref{sec: robustness}: {Statistical Robustness and Artifact Analysis. Statistical experiments demonstrating the robustness, evaluation stability, and validity of \textsc{MMTS-Bench} against dataset artifacts.}
    \item Appendix~\ref{sec:reproducibility statement}: Reproduction Statements.

\end{itemize}

\section{Details About MMTS-BENCH}
\label{sec:Details About MMTS-Bench}
\subsection{METRICS}
\label{sec:metrics}
We adopt a stratified evaluation scheme with specialized metrics tailored to each answer type within our answer space $\mathcal{A}$, ensuring comprehensive and fair assessment across diverse question formats.

\textbf{Categorical Evaluation ($\mathcal{A}_{mc}$ and $\mathcal{A}_{bf}$).} For both multiple-choice and binary-choice tasks, we utilize \textbf{Accuracy} as the primary evaluation metric, measuring exact match performance through the indicator function:
\begin{equation}
    \text{Accuracy} = \mathbb{I}(\hat{A} = A_{gt})
\end{equation}
where $\mathbb{I}(\hat{A} = A_{gt}) = 1$ if the predicted answer $\hat{A}$ matches ground-truth $A_{gt}$, and 0 otherwise.
    
\textbf{Numerical Evaluation ($\mathcal{A}_{num}$).} For numerical tasks, we employ two complementary metrics to capture both continuous proximity and threshold-based precision. 
Relative Accuracy quantifies the relative proximity between predicted and ground-truth values, yielding a normalized score in $[0, 1]$ where 1 indicates perfect prediction: 
\begin{equation}
    \text{Relative Accuracy} = \max\left(1.0 - \frac{|\hat{A} - A_{gt}|}{|A_{gt}|}, 0.0\right)
\end{equation}
Accuracy@10\% provides a stricter binary evaluation criterion, determining whether the prediction falls within a 10\% relative error tolerance:
\begin{equation}
        \text{Accuracy@10\%} = \mathbb{I}\left(\frac{|\hat{A} - A_{gt}|}{|A_{gt}|} \leq 0.1\right)
\end{equation}
The scoring function $\mathcal{M}(\hat{A}, A_{gt})$ introduced in our protocol is instantiated using these type-specific metrics, ensuring appropriate assessment based on answer type while maintaining consistency across the evaluation framework.

\subsection{Real-World Dataset Sources}
\label{sec:real-world sources}
The real-world component of our benchmark is constructed from the LOTSA \citep{woo2024pushing, godahewa2021monash, NEURIPS2023_ed73c36e, woo2024moirai} dataset collection. To ensure broad domain coverage while maintaining representativeness and high quality, we selected five major domains: Transport, Cloud Operations, Climate, Economics, and Healthcare. Representative datasets from these domains include Traffic Hourly, Alibaba Cluster Trace 2018, ERA5 2018, M4 Weekly, Hospital, and COVID deaths, with statistical parameters summarized in Table~\ref{tab:realworld-datasets}. Their details are as follows.

\textbf{Transport} (Traffic Hourly) The dataset originates from the California Department of Transportation. It records hourly highway occupancy rates from multiple sensors in the San Francisco Bay Area over a 48-month period (2015–2016). It contains 862 time series, each with 17,376 points in the range [0,1]. Because of the long time span and the strong seasonal patterns, we applied a sliding window with a maximum length of 672 points. To reduce token usage and irrelevant precision for LLMs, values were scaled by a factor of 100 and rounded to two decimal places.

\textbf{Cloud Operations} (Alibaba Cluster Trace 2018) This dataset describes CPU and memory utilization in a cluster of about 4,000 machines over eight days (from January 2 to January 8, 2018), sampled at five-minute intervals. It consists of 58,409 pairs of time series. Theoretical sequence length is 1,728 points, although some sequences are shorter due to missing samples (100–1,728 points). Values are within [0,100]. Because sequence lengths are moderate, no windowing was applied. Instead, we randomly sampled sequences and retained two decimal places.

\textbf{Climate} (ERA5 2018) The dataset comes from the European Centre for Medium-Range Weather Forecasts. It provides hourly global reanalysis data for 2018 at 2.8125$^\circ$ resolution (64$\times$128 grid points), covering 45 variables across seven pressure levels (50, 250, 500, 600, 700, 850, and 925 hPa). Each time series pair has 8,736 points. To construct our benchmark subset, we selected relative humidity and temperature from the seven pressure levels, with values within [0,100]. To capture spatial diversity, we randomly sampled 50 locations worldwide and then applied sliding windows of length 720. All values were rounded to two decimal places.

\textbf{Economics} (M4 Weekly) The dataset is a subset of the M4 Competition (2018), which consists of 100,000 time series across different frequencies. The weekly subset includes 359 economic and business-related series, such as sales, demand, and index values. Sequence lengths range from 80 to 2,597 points. To preserve potential seasonalities and balance sequence lengths, we used a sliding window with a maximum length of 104 points, approximately two years in length. Shorter series were kept in full. Values were rounded to two decimal places.

\textbf{Healthcare} (Hospital \& COVID deaths) The Hospital dataset records monthly patient counts related to medical products from January 2000 to December 2006. It contains 767 series of length 72. We applied sliding windows with common monthly cut lengths of 36, 60, and 72 points. All values were rounded to two decimal places.The COVID deaths dataset is sourced from the Johns Hopkins University repository. It contains cumulative daily death counts for countries and regions from January 22 to August 20, 2020. It consists of 266 daily series, each 182 points long. We applied a sliding window with a maximum length of 180 points. All values were rounded to two decimal places.


\begin{table}[htbp]
\centering
\caption{Statistical parameters of subsets in LOTSA.}
\label{tab:realworld-datasets}
\resizebox{\columnwidth}{!}{%
\begin{tabular}{lccccc}
\toprule
\textbf{Dataset} & \textbf{Domain} & \textbf{Frequency} & \textbf{\#Time Series} & \textbf{\#Obs.} & \textbf{\#Vars} \\
\midrule
Traffic Hourly & Transport & H & 862 & 14,978,112 & 1 \\
Alibaba Cluster Trace 2018 & CloudOps & 5T & 58,409 & 95,192,530 & 2 \\
ERA5 2018 & Climate & H & 245,760 & 2,146,959,000 & 45 \\
M4 Weekly & Economics & W & 359 & 366,912 & 1 \\
Hospital & Healthcare & M & 767 & 55,224 & 1 \\
COVID Deaths & Healthcare & D & 266 & 48,412 & 1 \\
\bottomrule
\end{tabular}%
}
\end{table}

\begin{table*}[htbp]
\centering
\caption{Core Dimensions of time series tasks described by \textsc{MMTS-Bench}, their covered subtask types, and related subsets.}
\label{tab:full_definition}
\resizebox{\textwidth}{!}{%
\renewcommand{\arraystretch}{1.1}
\setlength{\tabcolsep}{5pt}
\begin{tabular}{l l p{10cm} l}
\toprule
\textbf{Dimensions} & \textbf{Subtasks} & \textbf{Definition} & \textbf{Related Subsets} \\
\midrule
\multirow{3}{*}{Structural Awareness}
& Non-Stationarity & Analyzes statistical properties of concatenated subsequences. & \\
& Local-Global & Locates and analyzes specific sequence segments. & Base \\
& Uni.-Multivariate & Processes and analyzes multiple time series data jointly. & \\
\midrule
\multirow{5}{*}{Feature Analysis}
& Trend & Identifies long-term directional patterns and trend strength. & \\
& Seasonality & Captures seasonal patterns and seasonality strength. & \\
& Noise & Distinguishes random fluctuations from signal components. & Base, InWild \\
& Volatility & Quantifies temporal variability and instability. & \\
& Basic & Computes fundamental statistics (mean, variance, range, etc.). & \\
\midrule
\multirow{5}{*}{Temporal Reasoning}
& Deductive & Applies general rules to infer properties of specific intervals. & \\
& Inductive & Generalizes characteristics from observed sequences. & \\
& Causal & Identifies causal or lead-lag relationships between series. & InWild \\
& Analogical & Infers similarity by comparing temporal patterns. & \\
& Counterfactual & Predicts outcomes under hypothetical changes. & \\
\midrule
\multirow{4}{*}{Sequence Matching}
& Isomorphic & Finds the most similar sequence under equal-length constraints. & \\
& Robust & Robustly matches patterns under preprocessing transformations. & Match \\
& Positioning & Locates target patterns within longer sequences. & \\
& Reverse & Recognizes similarity under temporal reversal. & \\
\midrule
\multirow{2}{*}{Cross-Modal}
& TS→Semantic & Converts time series patterns into textual descriptions. & \\
& Semantic→TS & Maps textual descriptions to corresponding time series data. & Align \\
\bottomrule
\end{tabular}%
}
\end{table*}

\subsection{DATASETS CLASSIFICATION}
\label{sec:appendix DATASETS CLASSIFICATION}

To systematically evaluate LLMs' capabilities across the proposed multi-dimensional framework, we construct four specialized subsets that collectively form the core of the \textsc{MMTS-Bench}. The design of these subsets is based on a core hypothesis: a model's understanding of time series is hierarchical and progressive, where deficiencies in foundational analytical abilities lead to systematic biases in higher-level reasoning tasks. To operationalize this hypothesis, we decompose temporal understanding into five core dimensions.

We decompose time series understanding into several information dimensions, each representing a unique and non-substitutable processing requirement. Since a single task typically relies on only a subset of these dimensions, we organize and annotate our tasks accordingly to prevent mixed or redundant definitions. The notion of hierarchy means that higher-level capabilities are the result of combining multiple dimensions, rather than introducing entirely new dimensions. Although the dimensions in the Base and InWild subsets are related, we deliberately assign them distinct roles. The Base subset is designed to isolate and “activate” individual dimensions (e.g., structural awareness) in controlled settings. In contrast, InWild tasks, drawn from realistic scenarios, necessitate the simultaneous integration of multiple dimensions. By explicitly annotating these dimensional dependencies, we enable fine-grained failure analysis: poor performance on an InWild task can be traced to deficiencies in specific dimensions (e.g., structural awareness), rather than being attributed to an undifferentiated notion of overall failure.

To this end, we employ a dual-tier evaluation architecture: \textbf{foundational capability assessment} using synthetic data, followed by \textbf{advanced capability assessment} using real-world data. This approach ensures both precise, controlled evaluation of core competencies and a realistic assessment of practical performance.

\textbf{Base.} This subset is constructed using precisely controlled synthetic data, eliminating confounding variables present in real-world data, to provide standardized, fine-grained evaluation of a model's foundational time series analysis capabilities in a controlled environment. It contains 700 QA pairs that encompass two core dimensions of Structural Awareness ($D_s$) and Feature Analysis ($D_f$).

The following three subsets are constructed from real-world data in the LOTSA benchmark to evaluate advanced capabilities in complex scenarios.

\textbf{InWild.} This subset is constructed from across five specialized domains (Transport, CloudOps, Climate, Econ/Fin, Healthcare) to evaluate a model's capabilities in advanced time series understanding and reasoning. Through combinatorial arrangements of three core dimensions—Structural Awareness ($D_s$), Feature Analysis ($D_f$), and Temporal Reasoning ($D_r$)—it generates 1,084 QA pairs covering 140 subtask types.

\begin{figure*}[t]
\begin{center}
\includegraphics[width=0.8\textwidth]{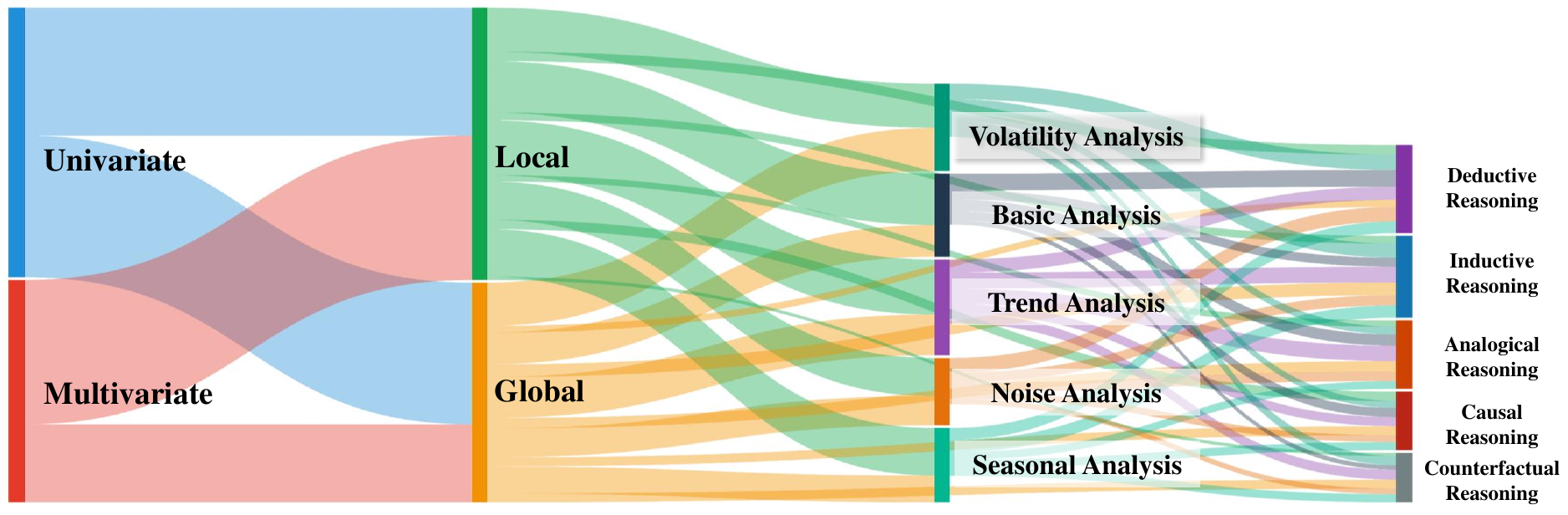}
\end{center}
\caption{Sankey Diagram of Subtask Labels in the InWild Subset. This diagram illustrates the relationships and transitions between subtask labels in the InWild subset, highlighting their interdependencies in a clear and intuitive way.}
\end{figure*}

\textbf{Match.} This subset is constructed from Transport and Econ/Fin domains to evaluate a model's performance in time series similarity matching and morphological correspondence. We generated four categories of sub-tasks by varying the Sequence Matching ($D_m$) dimensionality while holding Structural Awareness ($D_s$) dimensionality—stationarity, global scope, and univariate series—constant. The subset contains 400 QA pairs, with 100 pairs in each category.

\textbf{Align.} This subset is constructed from across five specialized domains (Transport, CloudOps, Climate) to systematically evaluate models' capabilities in bidirectional understanding and cross-modal conversion between numerical time series and natural language. From the perspective of the Cross-Modal Understanding dimension ($D_c$), we constructed 240 bidirectional QA pairs based on homologous time series sequences.

This hierarchical dataset architecture enables \textsc{MMTS-Bench} to provide comprehensive evaluation spanning from foundational analytical skills to advanced analytical capabilities, providing detailed diagnostic capability profiles that identify specific strengths and weaknesses of models in the context of time series understanding and reasoning tasks.

\begin{table*}[h!]
\centering
\caption[Horizontal comparison of existing datasets for time series understanding and reasoning.]{This table presents a horizontal comparison of existing datasets for time series understanding and reasoning. The \textbf{TS Type} column denotes the source of time series data, where “R” refers to real-world data and “S” refers to synthetic data. The \textbf{Domain} column indicates the application domain of the time series. \textbf{Question Type} specifies the types of questions included in the dataset, such as Choice or Numerical. \textbf{Taxonomy} shows how each dataset categorizes different capability dimensions. \textbf{Input Method} describes how time series are processed in these works, including TS-as-text, TS-encoded, and TS tokens. \textbf{Cross-Var.} indicate whether the dataset contains cross-variable analysis tasks, respectively. Finally, the \textbf{Generation} column outlines the construction methods of different datasets. For the detailed comparison table, please refer to \url{https://anonymous.4open.science/r/MMTS-BENCH-BEF7/comparison.md}}
\resizebox{\textwidth}{!}{
\begin{tabular}{l|cccccccccccc}
\toprule
\textbf{Dataset} & \textbf{TS Type} & \textbf{Domain} & \textbf{Size} & \textbf{Question Type} & \textbf{Taxonomy} & \textbf{Reasoning} & \textbf{Input Method} & \textbf{Cross-Var.} & \textbf{Generation} \\
\midrule
EngineMT-QA & R & Single & 11K/-- & Choice & Flat & \cmark & TS Encoded  & \xmark & Templates + Polishment \\
Time-MQA-TSQA & R & Multiple & 200K/1.4K & Choice+Num. & Flat & \cmark & TS-as-Text  & \xmark & Single-round prompting \\
ChatTime-TSQA & S & -- & 48K/0 & Choice & Flat & \xmark & TS tokens  & \xmark & Templates \\
TimeSeriesExam & S & -- & 746/-- & Choice & Flat & \cmark & --   & \cmark & Templates \\
ChatTS  & R+S & Multiple & 11K/525 & Choice+Num. & Flat & \cmark & TS Encoded  & \cmark & TS Self-Evol \\
Chat-TS & R & Multiple & 3741/100 & Choice & Flat & \cmark & TS tokens  & \xmark & Single-round prompting \\
MMTS-Bench & R+S & Multiple & 2524/1724 & Choice+Num. & Multi-level & \cmark & --  & \cmark & 3-stage prompting + Templates \\
\bottomrule
\end{tabular}
}
\end{table*}

\subsection{DATASET CONSTRUCTION METHOD}

To comprehensively evaluate multimodal time series understanding capabilities, we develop a dual-pathway construction methodology combining synthetic and real-world data sources. The synthetic pathway employs modular component synthesis with systematic parameter control to enable controlled evaluation of fundamental time series properties. The real-world pathway leverages progressive conversational frameworks, similarity matching algorithms, and cross-modal conversion techniques to construct comprehensive evaluation benchmarks. This approach yields four specialized subsets: \textbf{Base} for controlled synthetic evaluation, and \textbf{InWild} for multi-dimensional reasoning, \textbf{Match} for similarity matching, and \textbf{Align} for cross-modal understanding using authentic LOSTA data.

\subsubsection{SYNTHETIC DATASET}
\label{sec:appendix SYNTHETIC DATASET}


The time series data within the \textbf{Base} subset is generated through a modular synthesis approach. This process begins with the creation of three fundamental primitive components: (1) \textbf{Trend Components}, which can be configured with specific directions and magnitudes; (2) \textbf{Seasonal Components}, a diverse range of periodicities and waveforms are supported, encompassing both standard patterns (e.g., sine waves) and complex composite waveforms representative of real-world scenarios; and (3) \textbf{Noise Components}, where various noise types are included.

The final time series is formed by the superposition of these components. This principle is formally expressed through an additive model where the generated time series $y_t$ is the sum of three weighted components:
\begin{equation}
    \label{eq:decomposition}
    y^{(t)} = \tau_{\text{trend}}^{(t)} + \tau_{\text{seasonal}}^{(t)} + \tau_{\text{noise}}^{(t)}, \quad t=1,2,\dots,T.
\end{equation}
Each component $\tau^{(t)}$ is synthesized by scaling a corresponding \textbf{base signal} $S(t)$ with a randomly sampled weight $w$, such that $\tau_{\text{trend}}^{(t)} = w_{\text{trend}} \cdot S_{\text{trend}}(t)$, and similarly for the other components. The generation of these base signals is governed by a set of configurable parameters to ensure diversity, as detailed below.

First, a global \textbf{Sequence Length ($T$)} for all base signals is determined by sampling from the interval $[128, 2048]$ (corresponding to \texttt{min\_length} and \texttt{max\_length}). Then, each base signal is constructed as follows:

\begin{itemize}
    \item {Trend Base Signal ($S_{\text{trend}}^{(t)}$)} is designed to capture long-term, non-stationary behavior. It is modeled as an ARIMA(0,2,0) process, representing a second-order random walk:
\begin{equation}
    \label{eq:trend}
    S_{\text{trend}}^{(t)} = \sum_{n=1}^{t}\sum_{m=1}^{n} X_m, \quad \text{where} \quad X_t \sim \mathcal{N}(0,\sigma^2).
\end{equation}
For dataset generation, the smoothness of this trend is controlled by the second difference parameter, $\delta_s$ (\texttt{delta\_s}), which corresponds to the variance $\sigma^2$ and is sampled from $[0.01, 0.1]$. The overall amplitude of the base signal is sampled from $[0.1, 1000]$.

\item {Noise Base Signal ($S_{\text{noise}}^{(t)}$)} introduces various types of random fluctuations. It can be drawn from several distinct statistical distributions:
\begin{equation}
    \label{eq:noise}
    S_{\text{noise}}^{(t)} = X_t,
\end{equation}
where $X_t$ follows one of:
\begin{equation}
    \small
    X_t {=}\begin{cases}
        \mathcal{N}(\mu,\sigma^2), & \text{Gaussian} \\
        U(a,b), & \text{Uniform} \\
        \mathrm{round}(V_t/q)\!\cdot\! q,\; V_t {\sim} U(a,b), & \text{Quantization} \\
        X_{t-1}+V_t,\; V_t {\sim} U(a,b), & \text{Random walk}
    \end{cases}
\end{equation}


The specific parameters for each noise type (e.g., $\mu, \sigma^2$ for Gaussian, $a, b$ for Uniform) are configured to generate a variety of noise profiles. For the Quantization noise, the step size $q$ is specifically defined as one-tenth of the signal's amplitude.

    \item{Component Weights ($\boldsymbol{w}$)} After generating the three base signals, their respective contributions to the final time series are determined by a set of weights. The weights $\boldsymbol{w} = (w_{\text{trend}}, w_{\text{seasonal}}, w_{\text{noise}})$ are sampled from a symmetric Dirichlet distribution, $\boldsymbol{w} \sim \text{Dir}(\boldsymbol{\alpha})$ with $\boldsymbol{\alpha} = (1, 1, 1)$. This ensures an unbiased combination where the weights are positive and sum to one.

\end{itemize}

During the construction of the subset, the ground-truth labels for the four types of subtasks in the Feature Analysis ($D_f$) dimension are systematically generated based on the parameter system of the underlying primitive components. For the sub-tasks in the Structural Awareness ($D_s$) dimension, a differentiated construction strategy is adopted:

\begin{itemize}
    \item \textbf{Univariate vs. Multivariate:} Multivariate series are generated to evaluate a model's differential capabilities in analyzing univariate versus multivariate statistical properties.

    \item\textbf{Local vs. Global:} Descriptive and localization-based tasks are created by defining sub-intervals within the synthetic time series, thereby testing a model's local and global perceptual abilities.

    \item\textbf{Stationarity vs. Non-stationarity:} Composite series are constructed by concatenating two sub-series with distinct statistical properties. These are then used in conjunction with feature analysis tasks to assess a model's proficiency in identifying non-stationarity.
\end{itemize}


{QA pairs are systematically generated based on a set of 17 distinct templates. These templates cover a spectrum of tasks, ranging from qualitative feature analysis to quantitative numerical computation. Except for a subset of interval localization problems that require manual annotation, the vast majority of QA pairs are automatically generated. This template-based automation serves as the execution layer for our pipeline, the distinct advantages of which—specifically in terms of diversity and alignment precision compared to prior works—are detailed below.}

{\textbf{Advancements in Generation Pipeline.} While modular synthesis is a shared paradigm in recent works like TimeSeriesExam \citep{cai2024timeseriesexam} and ChatTS \citep{xie2024chatts}, our pipeline introduces critical enhancements in diversity, precision, and alignment.
\begin{itemize}
    \item \textbf{Generation Diversity:} Unlike TimeSeriesExam, which relies on simple base patterns (e.g., linear, exponential) and restricted sub-options, we adopt an STL-inspired decomposition into Trend, Seasonal, and Noise modules with rich control parameters. For instance, our trends are generated via second-order random walks (controlled by $\delta^2$ and initial values) rather than simple functions, and we incorporate diverse noise types (e.g., quantized, random-walk) and flexible combination strategies (Additive, Concatenated) that surpass the fixed combination schemes seen in ChatTS.
    \item \textbf{Precision and Alignment:} We log precise quantitative parameters (e.g., specific waveform configurations) rather than qualitative labels. This granular logging allows the QA generation engine to leverage exact numerical values, enabling the construction of nuanced evaluation items---such as differentiating between weak, medium, and strong seasonality based on the \texttt{seasonal\_strength} parameter---rather than limiting assessment to binary presence/absence questions.
\end{itemize}
}

\subsubsection{REAL-WORLD DATASET}
\label{sec: appendix REAL-WORLD DATASET}

To comprehensively evaluate multimodal time series understanding across diverse analytical dimensions, we construct three specialized subsets from real-world LOSTA data targeting distinct aspects of time series analysis: comprehensive QA-based evaluation (\textbf{InWild}), sequential similarity matching (\textbf{Match}), and cross-modal language understanding (\textbf{Align}). Each subset employs systematic construction pipelines with domain-specific sampling strategies and automated generation frameworks built upon authentic temporal data from five key domains.

\textbf{INWILD SUBSET}

This subset evaluates comprehensive understanding and reasoning capabilities in time series analysis through an innovative progressive, multi-turn conversational approach. The subset employs open-ended question templates to ensure rich diversity while avoiding rigid patterns. The generation process synergistically integrates three analytical dimensions: structural awareness, feature analysis, and temporal reasoning, creating 140 unique dimensional combinations that compel models to perform deep inference on underlying patterns and dynamic relationships.

\textbf{Three-Stage Progressive Generation Pipeline} Our dataset construction follows a systematic three-stage pipeline where each stage serves a distinct function in creating high-quality question-answering pairs.
\begin{itemize}
    \item \textbf{Stage 1: Initialization and Context Generation} The process begins with random sampling to define core parameters including question type (multiple-choice or true/false), task type (combining feature analysis and temporal reasoning capabilities), number of variables (univariate or multivariate), and analysis scope (local sub-sequence or global series). The system provides Claude 3.7 Sonnet with comprehensive multimodal context: time series visualizations, raw numerical sequences, domain metadata, and pre-computed statistical features from specialized libraries. The model generates a structured textual description tailored to the specified task type.

    \item \textbf{Stage 2: Task Specification and Reasoning Construction} Building upon the initial description, this stage further specifies the task according to predefined variables and analysis scope. For local-scope questions, the system provides indices for key time points. The model abstracts the structured information to construct specific questions, ground-truth answers, and detailed reasoning explanations.

    \item \textbf{Stage 3: Formatting and Quality Verification} The model formats outputs into standardized QA pairs according to designated question types while performing automated consistency verification across three dimensions: mathematical soundness of logic and calculations, descriptive accuracy between textual elements and data, and logical interpretability of reasoning coherence.
\end{itemize}

\textbf{MATCH SUBSET}

The \textbf{Match} subset is specifically designed to evaluate models' capabilities in time series similarity matching and morphological correspondence analysis. The subset employs standardized question-answer templates with fixed structural awareness parameters (stationarity, univariate, and global sequence) while systematically varying sequence matching dimensions to create four distinct matching paradigms.

\textbf{Four Matching Tasks}: (1) \textbf{Isomorphic Matching} evaluates models' fundamental ability to identify sequential similarity under identical temporal scales and data distributions, primarily assessing recognition accuracy of statistical characteristics and dynamic patterns; (2) \textbf{Robust Matching} evaluates models' resilience to maintain matching accuracy under data preprocessing transformations such as moving average smoothing, testing adaptability to data quality variations; (3) \textbf{Localization Matching} assesses models' precision in identifying target temporal patterns within extended time windows, focusing on temporal pattern retrieval and spatial localization performance; (4) \textbf{Reverse Matching} evaluates models' adaptability to temporal direction transformations, testing sequence correspondence recognition under time-reversed conditions.

These four progressive difficulty gradients comprehensively examine models' integrated capabilities in time series similarity measurement, pattern alignment, and morphological recognition under various constraints and challenging scenarios.

\textbf{Dataset Construction Pipeline} The construction process begins with segmenting sequences based on their typical periodicity patterns. Using a sliding window approach, we calculate Dynamic Time Warping (DTW) distances to identify four segments: minimum DTW distance, median distance, maximum distance, and second-maximum distance. We then randomly shuffle the subsequences of these four segments to serve as the four options in question–answer pairs. We construct the final dataset using standardized QA templates, covering four sequence matching tasks across two domains.

\textbf{ALIGN SUBSET}
\label{sec:appendix ALIGN SUBSET}

{The \textbf{Align} subset evaluates models' bidirectional conversion capabilities between time series data and natural language. It is intentionally designed as an alignment calibration subtask, aiming to assess whether a model can accurately match time series with natural language descriptions under conditions of low linguistic ambiguity. Consequently, the descriptions are intentionally specific regarding trend ranges and magnitude changes, ensuring the evaluation focuses on precise cross-modal alignment.}

\textbf{Bidirectional Cross-Modal Tasks} (1) \textbf{Time series to Semantic Conversion} requires models to identify which textual description best matches a given time series' statistical characteristics and dynamic patterns. (2) \textbf{Semantic to Time series Conversion} requires models to select the temporal sequence that best corresponds to a given natural language description of trends, fluctuations, and periodicity.


{\textbf{Dataset Construction Pipeline} The subset employs symmetric construction ensuring task consistency. The data originates from three real-world domains: Traffic, CloudOps, and Climate. To prevent shortcut learning based on domain or value range, we implemented two strategies: \textbf{(1) Value Scaling.} CloudOps and Climate series are scaled to similar ranges to mitigate magnitude-based biases. \textbf{(2) Controlled Sampling.} For each correct sample, we select three distractors. We ensure that at least one distractor comes from the same domain and shares a similar statistical distribution as the correct answer. This forces models to rely on fine-grained patterns and numeric details rather than domain priors.

For time series to semantic tasks, Claude 3.7 Sonnet generates feature descriptions for four temporal samples, which serve as answer options. For semantic to time series tasks, the model generates structured descriptions of correct samples, which become question prompts with temporal sequences as options.


\begin{figure*}[htbp]
\centering
\begin{subfigure}{\textwidth}
    \centering
    \includegraphics[width=\textwidth]{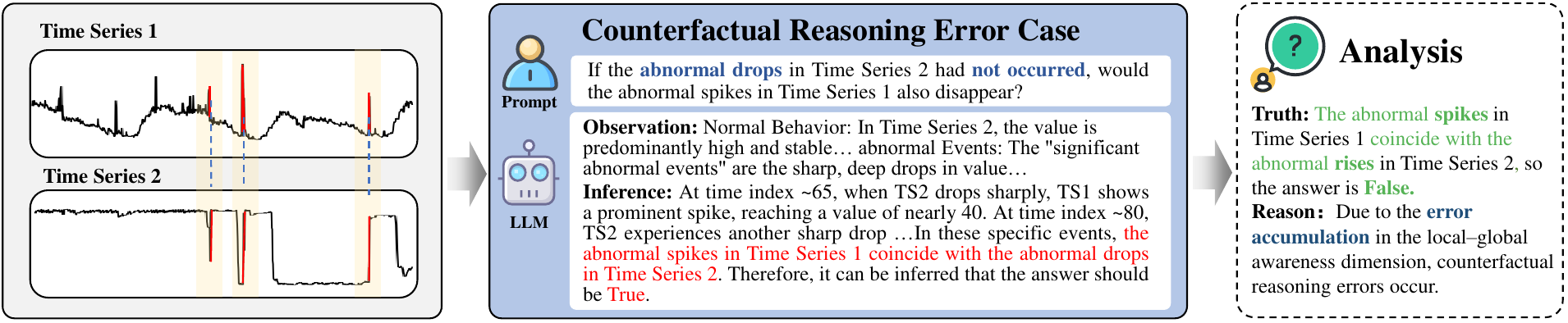}
    \caption{Counterfactual reasoning error case analysis}
\end{subfigure}

\begin{subfigure}{\textwidth}
    \centering
    \includegraphics[width=\textwidth]{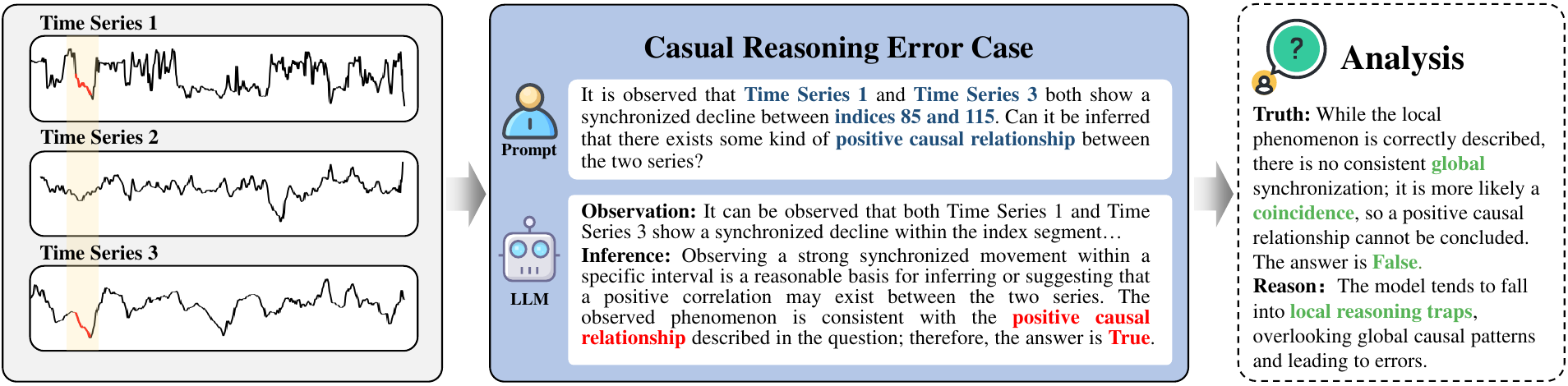}
    \caption{Casual reasoning error case analysis}
\end{subfigure}
\caption{The top subfigure illustrates a counterfactual reasoning error case, whereas the bottom subfigure illustrates a causal reasoning error case. In both subfigures, the left panel visualizes the time series data, the central panel presents the questions together with the LLM responses, and the right panel reports the ground-truth answers accompanied by error analysis. Red text highlights the erroneous reasoning points, green text denotes the correct reasoning points, and blue text marks the keywords from the questions.}
\label{fig:error_case}
\end{figure*}

\section{Evaluation and Experimental Setup}
\label{sec:Standardized TS Input Format}

We use the following prompts to standardize multiple-choice QA and numerical QA evaluation. The system prompt mandates the answer format and ambiguity policy; the user prompt injects per-item content. The model's output is scored by extracting the single letter or number inside the \texttt{<final\_answer>} tag.

\noindent\textbf{System Prompt for Multiple-Choice QA:}
\begin{lstlisting}[style=promptstyle, basicstyle=\ttfamily\footnotesize]
You are an expert AI assistant specialized in answering 
multiple-choice questions with high accuracy and 
consistency. Your task is to analyze questions carefully 
and provide clear, well-reasoned answers.

IMPORTANT INSTRUCTIONS:
1. You must select your answer from the given options only
2. Your final answer must be a single letter (A, B, C, 
   D, etc.)
3. If the question seems ambiguous, choose the most 
   reasonable interpretation
4. Do not make up information not provided in the question

RESPONSE FORMAT:
You must structure your final answer exactly as follows:
<final_answer>
[State your chosen option as a single letter: A, B, C, 
or D]
</final_answer>

Remember: Your final answer should contain ONLY the 
letter of your chosen option, nothing else.
\end{lstlisting}

\noindent\textbf{System Prompt for Numerical QA:}
\begin{lstlisting}[style=promptstyle, basicstyle=\ttfamily\footnotesize]
You are an expert AI assistant specialized in answering 
numerical questions with high accuracy and consistency. 
Your task is to analyze questions carefully and provide 
precise numerical answers.

IMPORTANT INSTRUCTIONS:
1. You must provide a numerical answer (integer, decimal, 
   or scientific notation)
2. Your final answer must be a single number only
3. If the question seems ambiguous, choose the most 
   reasonable interpretation
4. Do not include units unless specifically requested in 
   the question

RESPONSE FORMAT:
You must structure your final answer exactly as follows:
<final_answer>
[State your numerical answer as a single number only]
</final_answer>

Remember: Your final answer should contain ONLY the 
numerical value, nothing else.
\end{lstlisting}

\noindent\textbf{User Prompt:}
\begin{lstlisting}[style=promptstyle, basicstyle=\ttfamily\footnotesize]
Please answer the following multiple-choice/numerical 
question based on the given information:

Question: {question}
{given_values_str}
Available Options: {option}

Please analyze this question carefully, consider the 
given value and all available options, then provide your 
answer following the exact format specified in the system 
instructions.
\end{lstlisting}

\subsection{Time Series Visualization Setup}
We follow the plotting style of Zhuang et al.\ \cite{zhuang2024see} and adapt it for multi-channel time series. Specifically, we preserve the single-channel resolution (1500\,$\times$\,320 at 100\,dpi) and scale the figure height linearly with the number of channels by stacking channel-wise subplots with a fixed per-channel height (320\,px at 100\,dpi). This keeps a consistent time axis across channels while maintaining comparable vertical resolution per channel.

\begin{lstlisting}[style=pythonsnippet, basicstyle=\ttfamily\footnotesize]
def plot_time_series_as_image(value_list):
    if len(value_list) > 8:
        num_channels = 1           # single-channel
    else:
        num_channels = len(value_list)  # multi-channel

    width_inches = 15.0            # 1500 px / 100 dpi
    height_per_channel = 3.2       # 320 px / 100 dpi
    total_height = height_per_channel * num_channels
    fig = plt.figure(
        figsize=(width_inches, total_height), dpi=100
    )

    if num_channels == 1:
        plt.plot(range(len(value_list)),
                 value_list, 'b-', linewidth=1.5)
        plt.title('Time Series', fontsize=12)
        plt.xlabel('Time Index', fontsize=10)
        plt.ylabel('Value', fontsize=10)
        plt.grid(True, alpha=0.3)
        plt.xlim(0, len(value_list) - 1)
    else:
        for i, ch in enumerate(value_list):
            plt.subplot(num_channels, 1, i + 1)
            plt.plot(range(len(ch)),
                     ch, 'b-', linewidth=1.5)
            plt.title(f'Time Series {i+1}', fontsize=10)
            plt.xlabel('Time Index', fontsize=8)
            plt.ylabel('Value', fontsize=8)
            plt.grid(True, alpha=0.3)
            plt.xlim(0, len(ch) - 1)
    plt.tight_layout()
\end{lstlisting}

\subsection{Experimental Setups for Ablation Study}
\label{appen:ablation_study}

In ChatTS's\footnote{\url{https://huggingface.co/bytedance-research/ChatTS-14B}} original implementation, it employs {Qwen2.5-14B-Instruct} as the backbone LLM, with a 5-layer MLP as the time series encoder. During training\footnote{\url{https://github.com/xiezhe-24/ChatTS-Training}}, textual embeddings are aligned with time series embeddings to equip the model with time series reasoning capabilities. To examine the role of the encoder, we replaced the MLP with alternative architectures, including CNN and Transformer encoders with variable depth. We further tested the effect of learnable positional embeddings or index-based positional features introduced into the time series input.

Our experiments use {Qwen2.5-3B-Instruct} as the LLM backbone, and we also report its text-only baseline performance on \textsc{MMTS-Bench}. For comparison, we include performance of {Qwen2.5-14B-Instruct}, allowing us to isolate the effect of LLM backbone size. Finally, since ChatTS incorporates a prompt prefix that contains statistical information (e.g., offset, scale factor, length, max/min values, left/right boundary values), we tested models trained with and without this prefix to measure its contribution.

All models were evaluated on \textbf{InWild}, \textbf{Match}, and \textbf{Align}.

\clearpage
\section{Full Results}
\label{sec:appendix:Full Results}

\begin{table*}[bp]
\centering
\caption{The \textit{Accuracy} metric of different models on the \textbf{Base} subset's Choice split.
$^{cot}$ denotes \textit{thinking} mode.
$^1$ denotes models evaluated without any time series input.
$^2$ denotes ChatTS without built-in statistical computation module.
\texttt{-VL} = Vision-Language.
'TS' stands for Time Series modality, as time-series-specific models introduce a TS encoder.
'--' indicates that the model failed to respond correctly.
\textbf{\underline{Bold underlined}} values indicate the best performance within each category for each metric, and \textbf{bold} values indicate the second-best performance.
Stat. and Non-Stat. columns represent the questions with staionary and non-stationary time series, respectively. }
\label{tab:MMTS-Base-performance-qa}
\renewcommand{\arraystretch}{1.2} 
\setlength{\tabcolsep}{4pt}       
\resizebox{\textwidth}{!}{
\begin{tabular}{l|l c|c|ccc|cc}
\toprule
\textbf{Category} & \textbf{Model Name} & \textbf{Modality} & \textbf{Total} & \textbf{Trend} & \textbf{Seasonality} & \textbf{Noise} & \textbf{Local} & \textbf{Overall} \\
\midrule
\multirow{13}{*}{Open-source}
& DeepSeek-V3 & Text & 0.41 & 0.49 & 0.37 & 0.33 & \textbf{0.53} & 0.49 \\
& Kimi-K2 & Text & \textbf{\underline{0.45}} & \textbf{0.50} & 0.40 & \textbf{\underline{0.39}} & \textbf{\underline{0.55}} & \textbf{0.50} \\
& Qwen3-32b$^{cot}$ & Text & \textbf{0.42} & \textbf{\underline{0.53}} & \textbf{\underline{0.45}} & \textbf{0.37} & 0.32 & \textbf{\underline{0.53}} \\
& Qwen3-32b & Text & 0.40 & 0.46 & 0.37 & 0.35 & 0.47 & 0.46 \\
& Qwen3-8b$^{cot}$ & Text & 0.35 & 0.38 & 0.40 & 0.33 & 0.26 & 0.38 \\
& Qwen3-8b & Text & 0.32 & 0.33 & 0.27 & 0.26 & 0.50 & 0.33 \\
& Qwen2.5-32b & Text & 0.35 & 0.39 & 0.28 & 0.30 & 0.48 & 0.39 \\
& Qwen2.5-14b & Text & 0.35 & 0.39 & 0.31 & 0.27 & 0.46 & 0.39 \\
& Qwen2.5-7b & Text & 0.33 & 0.44 & 0.22 & 0.28 & 0.46 & 0.44 \\
& Qwen2.5-32b${^1}$ & Text & 0.30 & 0.39 & 0.23 & 0.19 & 0.44 & 0.39 \\
& Qwen2.5-7b-VL & Text & 0.29 & 0.35 & 0.23 & 0.24 & 0.36 & 0.35 \\
& Qwen2.5-7b-VL & Vision & 0.32 & 0.32 & \textbf{0.41} & 0.28 & 0.29 & 0.32 \\
& Qwen2.5-7b-VL & V+T & 0.34 & 0.41 & 0.34 & 0.27 & 0.34 & 0.41 \\
\midrule
\multirow{11}{*}{Closed-source}
& Claude-3.7-Sonnet & Text & 0.49 & \textbf{0.59} & 0.54 & 0.35 & 0.52 & \textbf{0.59 }\\
& Claude-Sonnet-4 & Text & 0.49 & 0.57 & 0.49 & 0.35 & \textbf{\underline{0.63}} & 0.57 \\
& Gemini-2.5-Pro & Text & 0.48 & 0.50 & 0.51 & 0.39 & 0.59 & 0.50 \\
& Gemini-2.5-Flash & Text & 0.39 & 0.49 & 0.36 & 0.24 & 0.56 & 0.49 \\
& GPT-5-Minimal & Text & 0.45 & 0.45 & 0.40 & \textbf{0.42} & 0.56 & 0.45 \\
& GPT-5-High & Text & \textbf{0.51} & 0.53 & 0.51 & \textbf{\underline{0.45}} & \textbf{0.60} & 0.53 \\
& GPT-4o & Text & 0.42 & 0.51 & 0.36 & 0.34 & 0.52 & 0.51 \\
& GPT-4o & Vision & \textbf{\underline{0.55}} & \textbf{\underline{0.60}} & \textbf{\underline{0.64}} & \textbf{0.42} & 0.56 & \textbf{\underline{0.60}} \\
& GPT-4o & V+T & \textbf{0.51} & 0.53 & \textbf{0.60} & 0.39 & 0.56 & 0.53 \\
& GPT-4o-mini & Text & 0.36 & 0.43 & 0.34 & 0.28 & 0.44 & 0.43 \\
\midrule
\multirow{4}{*}{Time-series}
& ChatTS & TS & 0.39 & 0.42 & 0.39 & 0.31 & 0.49 & 0.37  \\
& ChatTS$^{2}$ & TS & 0.39 & 0.37 & 0.39 & 0.34 & 0.53 & 0.36 \\
& ITFormer & TS & 0.31 & 0.30 & 0.29 & 0.28 & 0.42 & 0.29 \\
& ChatTime & TS & -- & -- & -- & -- & --  & --  \\
\bottomrule
\end{tabular}
}
\end{table*}

\begin{table*}[ht]
\centering
\caption{The \textit{Accuracy@10\%} metric of different models on the \textbf{Base} subset's numerical split.
$^{cot}$ denotes \textit{thinking} mode.
$^2$ denotes ChatTS without built-in statistical computation module.
 \texttt{-VL} = Vision-Language.
'TS' stands for Time Series modality, as time-series-specific models introduce a TS encoder.
'--' indicates that the model failed to respond correctly.
\textbf{\underline{Bold underlined}} values indicate the best performance within each category for each metric, and \textbf{bold} values indicate the second-best performance.
Stat. and Non-Stat. columns represent the questions with staionary and non-stationary time series, respectively. }
\label{tab:MMTS-Base-performance-1}
\renewcommand{\arraystretch}{1.2} 
\setlength{\tabcolsep}{4pt}       
\resizebox{\textwidth}{!}{
\begin{tabular}{l|l c|c|c c c|c c| c c | c c}
\toprule
\textbf{Category} & \textbf{Model Name} & \textbf{Modality} & \textbf{Total} & \textbf{Trend} & \textbf{Seasonality} & \textbf{Basic} & \textbf{Stat.} & \textbf{Non-Stat.} & \textbf{Local} & \textbf{Overall} & \textbf{Uni-Var.} & \textbf{Multi-Var.}\\
\midrule
\multirow{13}{*}{Open-source}
& DeepSeek-V3 & Text & \textbf{0.41} & \textbf{\underline{0.08}} & 0.02 & \textbf{\underline{0.57}} & \textbf{\underline{0.57}} & \textbf{0.52} & \textbf{0.21} & \textbf{\underline{0.47}} & \textbf{\underline{0.57}} & \textbf{\underline{0.56}} \\
& Kimi-K2 & Text & \textbf{\underline{0.42}} & 0.04 & \textbf{\underline{0.13}} & \textbf{0.56} & \textbf{0.56} & \textbf{\underline{0.56}} & 0.19 & \textbf{0.45} & \textbf{0.56} & \textbf{0.54} \\
& Qwen3-32b$^{cot}$ & Text & 0.31 & 0.00 & \textbf{0.10} & 0.40 & 0.40 & 0.40 & 0.20 & 0.32 & 0.40 & 0.33 \\
& Qwen3-32b & Text & 0.35 & 0.02 & 0.03 & 0.47 & 0.47 & 0.50 & 0.07 & 0.38 & 0.47 & 0.42 \\
& Qwen3-8b$^{cot}$ & Text & 0.27 & 0.00 & 0.00 & 0.34 & 0.34 & 0.38 & 0.13 & 0.27 & 0.34 & 0.29 \\
& Qwen3-8b & Text & 0.26 & 0.02 & 0.03 & 0.33 & 0.33 & 0.39 & 0.07 & 0.27 & 0.33 & 0.26 \\
& Qwen2.5-32b & Text & 0.34 & 0.03 & 0.00 & 0.45 & 0.45 & 0.50 & 0.11 & 0.36 & 0.45 & 0.44 \\
& Qwen2.5-14b & Text & 0.25 & 0.05 & 0.00 & 0.29 & 0.29 & 0.41 & 0.04 & 0.24 & 0.29 & 0.34 \\
& Qwen2.5-7b & Text & 0.18 & \textbf{0.06} & 0.00 & 0.19 & 0.19 & 0.31 & 0.01 & 0.16 & 0.19 & 0.31 \\
& Qwen2.5-7b-VL & Text & 0.14 & 0.03 & 0.03 & 0.14 & 0.14 & 0.23 & 0.04 & 0.12 & 0.14 & 0.21 \\
& Qwen2.5-7b-VL & Vision & 0.25 & 0.05 & 0.09 & 0.28 & 0.28 & 0.27 & \textbf{\underline{0.27}} & 0.24 & 0.28 & 0.35 \\
& Qwen2.5-7b-VL & V+T & 0.19 & 0.05 & 0.04 & 0.21 & 0.21 & 0.27 & 0.09 & 0.18 & 0.21 & 0.31 \\
\midrule
\multirow{11}{*}{Closed-source}
& Claude-3.7-Sonnet & Text & \textbf{0.54} & 0.13 & 0.20 & \textbf{0.66} & \textbf{0.66} & \textbf{0.65} & 0.44 & \textbf{0.55} & \textbf{0.66} & \textbf{0.66} \\
& Claude-Sonnet-4 & Text & 0.53 & 0.04 & 0.33 & 0.62 & 0.62 & 0.61 & \textbf{0.53} & 0.50 & 0.62 & 0.66 \\
& Gemini-2.5-Pro & Text & \textbf{\underline{0.63}} & 0.14 & \textbf{\underline{0.41}} & \textbf{\underline{0.72}} & \textbf{\underline{0.72}} & \textbf{\underline{0.69}} & \textbf{\underline{0.71}} & \textbf{\underline{0.60}} & \textbf{\underline{0.72}} & \textbf{\underline{0.72}} \\
& Gemini-2.5-Flash & Text & 0.43 & 0.04 & 0.20 & 0.56 & 0.56 & 0.55 & 0.20 & 0.45 & 0.56 & 0.40 \\
& GPT-5-Minimal & Text & 0.42 & 0.10 & 0.09 & 0.58 & 0.58 & 0.53 & 0.16 & 0.48 & 0.58 & 0.57 \\
& GPT-4o & Text & 0.41 & 0.06 & 0.09 & 0.53 & 0.53 & 0.58 & 0.13 & 0.43 & 0.53 & 0.51 \\
& GPT-4o & Vision & 0.28 & \textbf{0.17} & 0.10 & 0.34 & 0.34 & 0.29 & 0.31 & 0.31 & 0.34 & 0.34 \\
& GPT-4o & V+T & 0.51 & \textbf{\underline{0.42}} & \textbf{0.37} & 0.59 & 0.59 & 0.57 & 0.31 & 0.56 & 0.59 & 0.56 \\
& GPT-4o-mini & Text & 0.33 & 0.02 & 0.00 & 0.40 & 0.40 & 0.51 & 0.09 & 0.32 & 0.40 & 0.43 \\
\midrule
\multirow{4}{*}{Time-series}
& ChatTS & TS & 0.37 & 0.36 & 0.43 & 0.41 & 0.41 & 0.42 & 0.12 & 0.40 & 0.41 & 0.40  \\
& ChatTS$^{2}$ & TS & 0.01 & 0.00 & 0.45 & 0.01 & 0.01 & 0.01 & 0.11 & 0.01 & 0.01 & 0.02 \\
& ITFormer & TS & 0.00 & 0.00 & 0.00 & 0.00 & 0.00 & 0.01 & 0.02 & 0.00 & 0.00 & 0.00 \\
& ChatTime & TS & -- & -- & -- & -- & --  & -- & -- & -- & -- & -- \\
\bottomrule
\end{tabular}
}
\end{table*}


\begin{figure*}[htbp]
\begin{center}
\includegraphics[width=\textwidth]{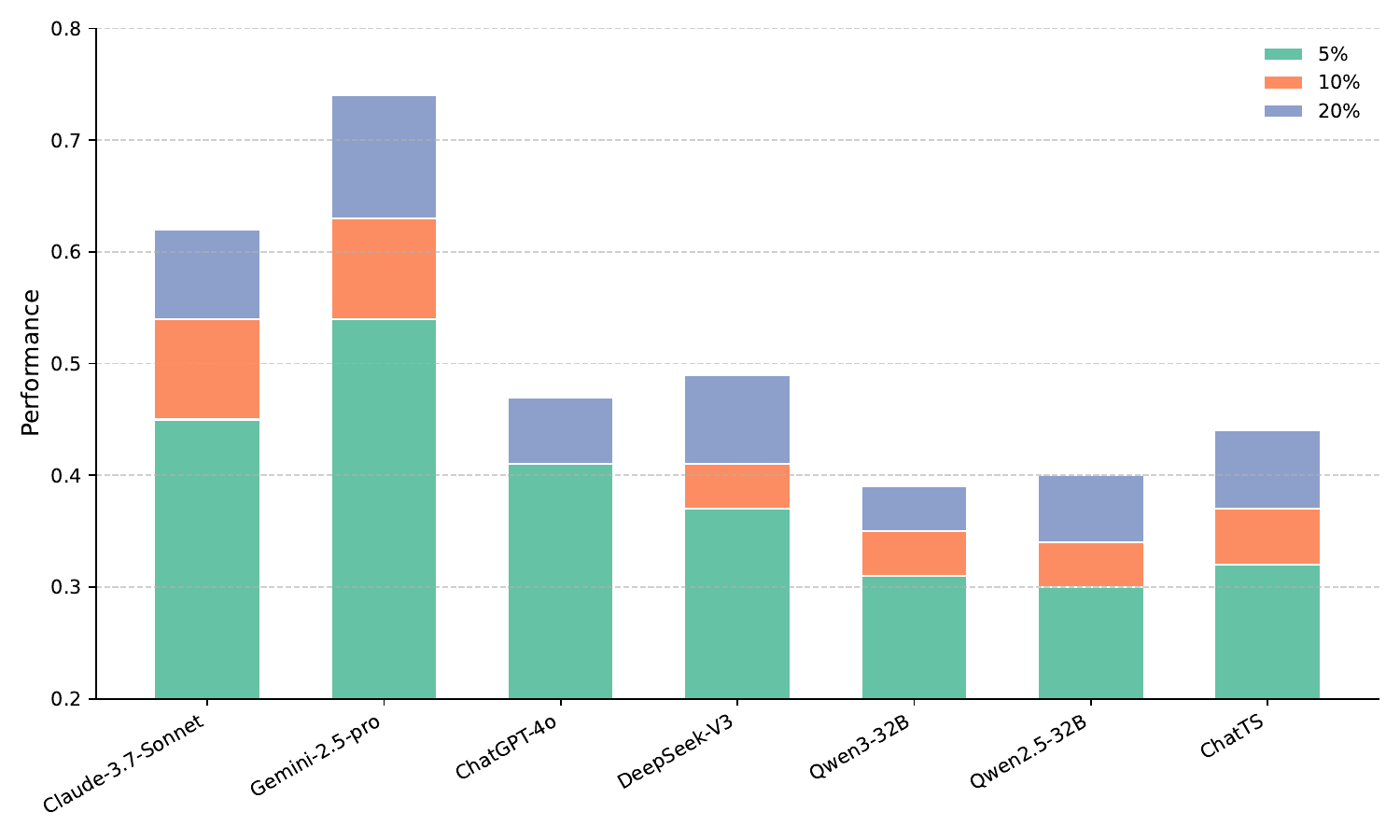}
\end{center}
\caption{This stacked bar chart illustrates the \textit{Accuracy@N\%} performance of several representative models (closed-source, open-source, and TS-LLMs) on the numerical split of the \textbf{Base} subset. To explore the hierarchical distribution of numerical reasoning ability, we report results for \texttt{N = 5, 10, 20}.
}
\label{fig:Acc_at_N}
\end{figure*}

\begin{table*}[ht]
\centering
\caption{The \textit{Relative Accuracy} metric of different models on the \textbf{Base} subset's numerical split.
$^{cot}$ denotes \textit{thinking} mode.
$^2$ denotes ChatTS without built-in statistical computation module.
 \texttt{-VL} = Vision-Language.
'TS' stands for Time Series modality, as time-series-specific models introduce a TS encoder.
'--' indicates that the model failed to respond correctly.
\textbf{\underline{Bold underlined}} values indicate the best performance within each category for each metric, and \textbf{bold} values indicate the second-best performance.
Stat. and Non-Stat. columns represent the questions with staionary and non-stationary time series, respectively. }
\label{tab:MMTS-Base-performance-2}
\renewcommand{\arraystretch}{1.2} 
\setlength{\tabcolsep}{4pt}       
\resizebox{\textwidth}{!}{
\begin{tabular}{l|l c|c|c c c|c c| c c | c c}
\toprule
\textbf{Category} & \textbf{Model Name} & \textbf{Modality} & \textbf{Total} & \textbf{Trend} & \textbf{Seasonality} & \textbf{Basic} & \textbf{Stat.} & \textbf{Non-Stat.} & \textbf{Local} & \textbf{Overall} & \textbf{Uni-Var.} & \textbf{Multi-Var.}\\
\midrule
\multirow{13}{*}{Open-source}
& DeepSeek-V3 & Text & \textbf{0.59} & \textbf{0.27} & 0.32 & \textbf{0.73} & \textbf{0.73} & \textbf{0.66} & \textbf{0.41} & \textbf{0.63} & \textbf{0.73} & \textbf{\underline{0.76}} \\
& Kimi-K2 & Text & \textbf{\underline{0.65}} & \textbf{\underline{0.37}} & \textbf{\underline{0.49}} & \textbf{\underline{0.74}} & \textbf{\underline{0.74}} & \textbf{\underline{0.71}} & \textbf{\underline{0.56}} & \textbf{\underline{0.66}} & \textbf{\underline{0.74}} & \textbf{0.72} \\
& Qwen3-32b$^{cot}$ & Text & 0.42 & 0.00 & 0.30 & 0.51 & 0.51 & 0.52 & 0.27 & 0.41 & 0.51 & 0.47 \\
& Qwen3-32b & Text & 0.49 & 0.07 & 0.25 & 0.61 & 0.61 & 0.64 & 0.29 & 0.50 & 0.61 & 0.61 \\
& Qwen3-8b$^{cot}$ & Text & 0.37 & 0.00 & 0.19 & 0.47 & 0.47 & 0.49 & 0.19 & 0.37 & 0.47 & 0.43 \\
& Qwen3-8b & Text & 0.40 & 0.13 & 0.19 & 0.49 & 0.49 & 0.54 & 0.16 & 0.42 & 0.49 & 0.46 \\
& Qwen2.5-32b & Text & 0.51 & 0.16 & 0.13 & 0.66 & 0.66 & 0.63 & 0.30 & 0.56 & 0.66 & 0.64 \\
& Qwen2.5-14b & Text & 0.44 & 0.14 & 0.10 & 0.51 & 0.51 & 0.59 & 0.29 & 0.43 & 0.51 & 0.55 \\
& Qwen2.5-7b & Text & 0.35 & 0.20 & 0.12 & 0.40 & 0.40 & 0.49 & 0.14 & 0.36 & 0.40 & 0.50 \\
& Qwen2.5-7b-VL & Text & 0.32 & 0.16 & 0.15 & 0.37 & 0.37 & 0.43 & 0.12 & 0.33 & 0.37 & 0.42 \\
& Qwen2.5-7b-VL & Vision & 0.40 & 0.26 & \textbf{0.36} & 0.44 & 0.44 & 0.48 & 0.22 & 0.4 & 0.44 & 0.50 \\
& Qwen2.5-7b-VL & V+T & 0.19 & 0.05 & 0.04 & 0.21 & 0.21 & 0.27 & 0.09 & 0.18 & 0.21 & 0.31 \\
\midrule
\multirow{11}{*}{Closed-source}
& Claude-3.7-Sonnet & Text & \textbf{0.74} & 0.51 & 0.59 & \textbf{0.80} & \textbf{0.80} & \textbf{0.79} & 0.72 & \textbf{0.74} & \textbf{0.80} & \textbf{0.80} \\
& Claude-Sonnet-4 & Text & 0.73 & 0.36 & 0.61 & 0.78 & 0.78 & 0.79 & \textbf{0.79} & 0.69 & 0.78 & \textbf{0.80} \\
& Gemini-2.5-Pro & Text & \textbf{\underline{0.82}} & \textbf{0.56} & \textbf{\underline{0.78}} & \textbf{\underline{0.85}} & \textbf{\underline{0.85}} & \textbf{\underline{0.83}} & \textbf{\underline{0.88}} & \textbf{\underline{0.79}} & \textbf{\underline{0.85}} & \textbf{\underline{0.83}} \\
& Gemini-2.5-Flash & Text & 0.57 & 0.10 & 0.55 & 0.69 & 0.69 & 0.66 & 0.34 & 0.57 & 0.69 & 0.53 \\
& GPT-5-Minimal & Text & 0.61 & 0.24 & 0.33 & 0.77 & 0.77 & 0.68 & 0.45 & 0.66 & 0.77 & 0.78 \\
& GPT-4o & Text & 0.60 & 0.32 & 0.35 & 0.73 & 0.73 & 0.7 & 0.38 & 0.64 & 0.73 & 0.67 \\
& GPT-4o & Vision & 0.58 & 0.30 & 0.71 & 0.60 & 0.60 & 0.55 & 0.63 & 0.55 & 0.60 & 0.57 \\
& GPT-4o & V+T & 0.70 & \textbf{\underline{0.59}} & \textbf{0.75} & 0.76 & 0.76 & 0.70 & 0.60 & 0.73 & 0.76 & 0.74 \\
& GPT-4o-mini & Text & 0.50 & 0.17 & 0.14 & 0.60 & 0.60 & 0.66 & 0.31 & 0.51 & 0.60 & 0.61 \\
\midrule
\multirow{4}{*}{Time-series}
& ChatTS & TS & 0.55 & 0.46 & 0.79 & 0.59 & 0.59 & 0.56 & 0.34 & 0.56 & 0.59 & 0.56  \\
& ChatTS$^{2}$ & TS & 0.19 & 0.06 & 0.76 & 0.10 & 0.10 & 0.10 & 0.36 & 0.09 & 0.10 & 0.10 \\
& ITFormer & TS & 0.15 & 0.02 & 0.06 & 0.15 & 0.15 & 0.19 & 0.23 & 0.12 & 0.15 & 0.18 \\
& ChatTime & TS & -- & -- & -- & -- & --  & -- & -- & -- & -- & -- \\
\bottomrule
\end{tabular}
}
\end{table*}

\begin{table*}[ht]
\centering
\caption{The \textit{Average Offset} metric of different models on the \textbf{Base} subset's numerical split.
$^{cot}$ denotes \textit{thinking} mode.
$^2$ denotes ChatTS without built-in statistical computation module.
 \texttt{-VL} = Vision-Language.
'TS' stands for Time Series modality, as time-series-specific models introduce a TS encoder.
'--' indicates that the model failed to respond correctly.
\textbf{\underline{Bold underlined}} values indicate the best performance within each category for each metric, and \textbf{bold} values indicate the second-best performance.
Stat. and Non-Stat. columns represent the questions with staionary and non-stationary time series, respectively. \textbf{H} remark means the average offset value is higher than \texttt{1e5}.}
\label{tab:MMTS-Base-performance-3}
\renewcommand{\arraystretch}{1.2} 
\setlength{\tabcolsep}{4pt}       
\resizebox{\textwidth}{!}{
\begin{tabular}{l|l c|c|c c c|c c| c c | c c}
\toprule
\textbf{Category} & \textbf{Model Name} & \textbf{Modality} & \textbf{Total} & \textbf{Trend} & \textbf{Seasonality} & \textbf{Basic} & \textbf{Stat.} & \textbf{Non-Stat.} & \textbf{Local} & \textbf{Overall} & \textbf{Uni-Var.} & \textbf{Multi-Var.}\\
\midrule
\multirow{13}{*}{Open-source}
& DeepSeek-V3 & Text & \textbf{\underline{1.06}} & \textbf{\underline{1.17}} & 0.68 & \textbf{\underline{0.69}} & \textbf{\underline{0.69}} & 1.64 & \textbf{0.69} & \textbf{\underline{0.79}} & \textbf{\underline{0.69}} & \textbf{1.94} \\
& Kimi-K2 & Text & \textbf{1.24} & \textbf{1.80} & \textbf{\underline{0.57}} & \textbf{1.14} & \textbf{1.14} & 1.68 & \textbf{\underline{0.49}} & \textbf{1.27} & \textbf{1.14} & \textbf{\underline{1.59}} \\
& Qwen3-32b$^{cot}$ & Text & 49.19 & 512.64 & 1.07 & 2.32 & 2.32 & 3.73 & 8.45 & 107.63 & 2.32 & 4.94 \\
& Qwen3-32b & Text & 4.57 & 34.94 & 0.75 & 1.43 & 1.43 & 1.48 & 2.72 & 8.34 & 1.43 & 3.17 \\
& Qwen3-8b$^{cot}$ & Text & 54.04 & 568.67 & 1.04 & 2.45 & 2.45 & 3.47 & 7.73 & 119.29 & 2.45 & 5.25 \\
& Qwen3-8b & Text & 6.01 & 39.83 & 0.81 & 3.05 & 3.05 & 1.76 & 4.96 & 10.64 & 3.05 & 4.15 \\
& Qwen2.5-32b & Text & 1.65 & 5.91 & 0.92 & 1.49 & 1.49 & \textbf{\underline{1.05}} & 1.22 & 2.40 & 1.49 & 3.77 \\
& Qwen2.5-14b & Text & 16.56 & 170.23 & 0.90 & 2.06 & 2.06 & \textbf{1.12} & 1.33 & 36.76 & 2.06 & 5.37 \\
& Qwen2.5-7b & Text & 3.02 & 3.49 & 0.88 & 3.20 & 3.20 & 1.97 & 6.00 & 3.26 & 3.20 & 4.81 \\
& Qwen2.5-7b-VL & Text & H & 7.06 & 1.82 & 5.55 & 5.55 & H & 98.73 & 5.86 & 5.55 & 15.49 \\
& Qwen2.5-7b-VL & Vision & H & 13.24 & \textbf{0.61} & 31.51 & 31.51 & 1.32 & 1.17 & 28.08 & 31.51 & H \\
& Qwen2.5-7b-VL & V+T & H & 34.87 & 1.01 & H & H & H & 17.56 & H & H & 4.15 \\
\midrule
\multirow{11}{*}{Closed-source}
& Claude-3.7-Sonnet & Text & \textbf{\underline{0.48}} & \textbf{\underline{1.24}} & 0.46 & \textbf{0.32} & \textbf{0.32} & \textbf{0.39} & 0.62 & \textbf{\underline{0.51}} & \textbf{0.32} & 0.50 \\
& Claude-Sonnet-4 & Text & \textbf{0.72} & \textbf{1.26} & 0.56 & 0.62 & 0.62 & 0.78 & \textbf{0.58} & 0.75 & 0.62 & \textbf{0.44} \\
& Gemini-2.5-Pro & Text & 1.13 & 9.62 & \textbf{0.27} & 0.36 & 0.36 & \textbf{\underline{0.30}} & \textbf{\underline{0.35}} & 2.21 & 0.36 & \textbf{\underline{0.27}} \\
& Gemini-2.5-Flash & Text & 34.04 & 350.9 & 0.62 & 2.49 & 2.49 & 1.48 & 8.87 & 74.38 & 2.49 & 5.57 \\
& GPT-5-Minimal & Text & 0.91 & 1.85 & 0.69 & 0.70 & 0.70 & 0.97 & 0.83 & 0.94 & 0.70 & 0.67 \\
& GPT-4o & Text & 2.27 & 16.22 & 0.71 & 0.45 & 0.45 & 1.18 & 1.40 & 3.70 & 0.45 & 2.57 \\
& GPT-4o & Vision & 1.47 & 2.70 & 0.30 & 2.46 & 2.46 & 1.10 & 0.63 & 2.50 & 2.46 & 4.18 \\
& GPT-4o & V+T & \textbf{0.72} & 1.66 & \textbf{\underline{0.25}} & \textbf{\underline{0.31}} & \textbf{\underline{0.31}} & 0.73 & 1.35 & \textbf{0.59} & \textbf{\underline{0.31}} & 0.48 \\
& GPT-4o-mini & Text & 1.24 & 2.58 & 0.86 & 0.94 & 0.94 & 1.30 & 1.21 & 1.28 & 0.94 & 1.63 \\
\midrule
\multirow{4}{*}{Time-series}
& ChatTS & TS & 1.79 & 6.75 & 0.23 & 1.68 & 1.68 & 1.34 & 0.90 & 2.72 & 1.68 & 2.26  \\
& ChatTS$^{2}$ & TS & 21.89 & 130.25 & 0.24 & 20.54 & 20.54 & 8.88 & 0.88 & 43.18 & 20.54 & 7.21 \\
& ITFormer & TS & H & 7.08 & 0.94 & 10.89 & 10.89 & 1.59 & 0.79 & 10.10 & 10.89 & H \\
& ChatTime & TS & -- & -- & -- & -- & --  & -- & -- & -- & -- & -- \\
\bottomrule
\end{tabular}
}
\end{table*}

\begin{table*}[ht]
\centering
\caption{Performance of different models on the \textbf{InWild} subset.
$^{cot}$ denotes \textit{thinking} mode.
$^1$ denotes models evaluated without any time series input.
$^2$ denotes ChatTS without built-in statistical computation module.
\texttt{-VL} = Vision-Language.
'TS' stands for Time Series modality, as time-series-specific models introduce a TS encoder.
'--' indicates that the model failed to respond correctly.
In reasoning tasks, abbreviations are: Ded. (Deductive), Ind. (Inductive), Analog. (Analogical), and Count. (Counterfactual).
Best and second-best results within each category are \textbf{\underline{underlined}} and \textbf{bolded}, respectively.}
\label{tab:MMTS-InWild-performance}
\renewcommand{\arraystretch}{1.2} 
\setlength{\tabcolsep}{3pt}       
\resizebox{\textwidth}{!}{
\begin{tabular}{l|l c|c|c|ccccc|c|ccccc}
\toprule
\multirow{2}{*}{\textbf{Category}} &
\multirow{2}{*}{\textbf{Model Name}} &
\multirow{2}{*}{\textbf{Modality}} &
\multirow{2}{*}{\textbf{Average}} &
\multicolumn{6}{c|}{\textbf{Feature Analysis}} &
\multicolumn{6}{c}{\textbf{Temporal Reasoning}} \\
\cmidrule(lr){5-10} \cmidrule(lr){11-16}
& & & & Acc. & Trend & Season & Noise & Volat. & Basic
  & Acc. & Ded. & Ind. & Causal & Analog. & Count. \\
\midrule
\multirow{13}{*}{Open-source}
& DeepSeek-V3 & Text & \textbf{0.61} & \textbf{0.67} & \textbf{\underline{0.73}} & \textbf{0.57} & 0.79 & \textbf{0.55} & \textbf{\underline{0.75}} & \textbf{0.59} & \textbf{0.60} & \textbf{0.63} & \textbf{0.52} & 0.58 & \textbf{\underline{0.57}} \\
& Kimi-K2 & Text & \textbf{\underline{0.63}} & \textbf{\underline{0.69}} & 0.71 & \textbf{\underline{0.63}} & \textbf{\underline{0.85}} & \textbf{\underline{0.58}} & \textbf{0.71} & \textbf{\underline{0.61}} & \textbf{\underline{0.62}} & \textbf{\underline{0.65}} & \textbf{\underline{0.56}} & \textbf{\underline{0.64}} & \textbf{0.55} \\
& Qwen3-32b$^{cot}$ & Text & 0.58 & 0.65 & \textbf{0.72} & 0.54 & \textbf{0.81} & \textbf{\underline{0.58}} & 0.65 & 0.55 & 0.52 & 0.60 & 0.49 & \textbf{0.61} & 0.48 \\
& Qwen3-32b & Text & 0.50 & 0.56 & 0.67 & 0.49 & 0.54 & 0.43 & 0.62 & 0.48 & 0.48 & 0.51 & 0.42 & 0.54 & 0.42 \\
& Qwen3-8b$^{cot}$ & Text & 0.50 & 0.57 & 0.54 & 0.51 & 0.74 & 0.54 & 0.55 & 0.47 & 0.41 & 0.52 & 0.43 & 0.49 & 0.48 \\
& Qwen3-8b & Text & 0.45 & 0.48 & 0.45 & 0.43 & 0.65 & 0.34 & 0.58 & 0.44 & 0.48 & 0.46 & 0.42 & 0.45 & 0.37 \\
& Qwen2.5-32b & Text & 0.53 & 0.62 & 0.66 & 0.53 & 0.77 & 0.52 & 0.62 & 0.49 & 0.47 & 0.53 & 0.51 & 0.50 & 0.44 \\
& Qwen2.5-14b & Text & 0.53 & 0.61 & 0.60 & 0.55 & 0.73 & \textbf{\underline{0.58}} & 0.63 & 0.49 & 0.48 & 0.53 & 0.48 & 0.53 & 0.41 \\
& Qwen2.5-7b & Text & 0.44 & 0.45 & 0.61 & 0.48 & 0.35 & 0.35 & 0.42 & 0.44 & 0.41 & 0.47 & 0.43 & 0.45 & 0.44 \\
& Qwen2.5-7b-VL & Text & 0.37 & 0.37 & 0.41 & 0.32 & 0.44 & 0.28 & 0.38 & 0.37 & 0.38 & 0.35 & 0.40 & 0.35 & 0.36 \\
& Qwen2.5-7b-VL & Vision & 0.36 & 0.37 & 0.39 & 0.44 & 0.37 & 0.30 & 0.33 & 0.35 & 0.38 & 0.36 & 0.36 & 0.33 & 0.33 \\
& Qwen2.5-7b-VL & V+T & 0.39 & 0.41 & 0.48 & 0.41 & 0.45 & 0.30 & 0.40 & 0.38 & 0.39 & 0.37 & 0.43 & 0.36 & 0.35 \\
& Qwen2.5-32b$^{1}$ & Text & 0.34 & 0.34 & 0.40 & 0.40 & 0.24 & 0.33 & 0.31 & 0.34 & 0.35 & 0.35 & 0.38 & 0.31 & 0.30 \\
\midrule
\multirow{18}{*}{Closed-source}
& Claude-3.7-Sonnet & Text & 0.69 & 0.78 & 0.81 & \textbf{\underline{0.72}} & 0.85 & 0.73 & 0.81 & 0.65 & 0.66 & 0.67 & 0.58 & 0.71 & 0.60 \\
& Claude-3.7-Sonnet & Vision & 0.69 & 0.75 & 0.82 & 0.67 & 0.82 & 0.70 & 0.75 & 0.66 & 0.64 & 0.72 & 0.62 & 0.71 & 0.60 \\
& Claude-3.7-Sonnet & V+T & \textbf{0.73} & 0.79 & \textbf{0.84} & \textbf{0.71} & 0.87 & 0.70 & 0.85 & \textbf{0.71} & 0.66 & \textbf{0.76} & \textbf{\underline{0.70}} & 0.74 & 0.66 \\
& Claude-Sonnet-4 & Text & 0.71 & 0.79 & 0.78 & 0.67 & \textbf{\underline{0.91}} & 0.70 & \textbf{0.89} & 0.68 & 0.67 & 0.75 & 0.60 & 0.69 & 0.63 \\
& Claude-Sonnet-4 & Vision & 0.67 & 0.74 & 0.84 & 0.59 & 0.87 & 0.63 & 0.79 & 0.64 & 0.59 & 0.75 & 0.55 & 0.66 & 0.60 \\
& Claude-Sonnet-4 & V+T & 0.71 & 0.78 & 0.81 & 0.64 & \textbf{0.90} & 0.72 & 0.86 & 0.68 & 0.65 & \textbf{0.76} & 0.62 & 0.72 & 0.66 \\
& Gemini-2.5-Pro & Text & 0.68 & 0.73 & 0.76 & 0.69 & 0.83 & 0.66 & 0.72 & 0.66 & 0.65 & \textbf{0.76} & 0.56 & 0.68 & 0.63 \\
& Gemini-2.5-Pro & Vision & 0.72 & \textbf{0.80} & \textbf{0.84} & 0.63 & 0.89 & \textbf{\underline{0.77}} & 0.85 & 0.69 & \textbf{0.69} & \textbf{0.76} & 0.60 & 0.71 & 0.64 \\
& Gemini-2.5-Pro & V+T & \textbf{\underline{0.76}} & \textbf{\underline{0.83}} & \textbf{\underline{0.86}} & \textbf{0.71} & 0.87 & \textbf{0.76} & \textbf{\underline{0.92}} & \textbf{\underline{0.73}} & \textbf{\underline{0.75}} & \textbf{\underline{0.78}} & 0.62 & \textbf{\underline{0.77}} & \textbf{\underline{0.69}} \\
& Gemini-2.5-Flash & Text & 0.50 & 0.50 & 0.53 & 0.40 & 0.59 & 0.45 & 0.55 & 0.51 & 0.47 & 0.57 & 0.45 & 0.56 & 0.45 \\
& GPT-5-High & Text & 0.72 & 0.76 & 0.82 & 0.67 & 0.87 & 0.60 & 0.85 & 0.70 & 0.69 & 0.72 & \textbf{0.66} & \textbf{0.76} & \textbf{0.68} \\
& GPT-5-Minimal & Text & 0.56 & 0.64 & 0.70 & 0.63 & 0.72 & 0.55 & 0.61 & 0.52 & 0.53 & 0.56 & 0.51 & 0.53 & 0.45 \\
& GPT-4.1 & Text & 0.58 & 0.64 & 0.63 & 0.57 & 0.76 & 0.58 & 0.68 & 0.56 & 0.58 & 0.61 & 0.52 & 0.52 & 0.55 \\
& GPT-4.1 & Vision & 0.58 & 0.61 & 0.70 & 0.46 & 0.70 & 0.51 & 0.67 & 0.57 & 0.57 & 0.61 & 0.57 & 0.55 & 0.56 \\
& GPT-4.1 & V+T & 0.63 & 0.67 & 0.73 & 0.55 & 0.85 & 0.60 & 0.68 & 0.62 & 0.62 & 0.69 & 0.59 & 0.59 & 0.56 \\
& GPT-4.1-Mini & Text & 0.59 & 0.66 & 0.69 & 0.49 & 0.88 & 0.55 & 0.73 & 0.56 & 0.56 & 0.60 & 0.54 & 0.56 & 0.49 \\
& GPT-4.1-Mini & Vision & 0.49 & 0.51 & 0.56 & 0.37 & 0.54 & 0.51 & 0.55 & 0.49 & 0.51 & 0.45 & 0.52 & 0.50 & 0.48 \\
& GPT-4.1-Mini & V+T & 0.58 & 0.64 & 0.68 & 0.54 & 0.75 & 0.53 & 0.73 & 0.56 & 0.59 & 0.61 & 0.55 & 0.54 & 0.44 \\
& GPT-4o & Text & 0.62 & 0.70 & 0.74 & 0.61 & 0.79 & 0.61 & 0.75 & 0.58 & 0.59 & 0.62 & 0.50 & 0.66 & 0.50 \\
& GPT-4o & Vision & 0.59 & 0.67 & 0.76 & 0.56 & 0.74 & 0.64 & 0.66 & 0.56 & 0.55 & 0.62 & 0.51 & 0.57 & 0.50 \\
& GPT-4o & V+T & 0.63 & 0.71 & 0.75 & 0.59 & 0.85 & 0.61 & 0.77 & 0.59 & 0.59 & 0.65 & 0.55 & 0.63 & 0.53 \\
\midrule
\multirow{4}{*}{Time-series}
& ChatTS & TS & 0.50 & 0.55 & 0.50 & 0.61 & 0.53 & 0.54 & 0.58 & 0.48 & 0.51 & 0.50 & 0.49 & 0.44 & 0.45 \\
& ChatTS$^{2}$ & TS & 0.48 & 0.51 & 0.53 & 0.55 & 0.48 & 0.52 & 0.50 & 0.48 & 0.50 & 0.50 & 0.52 & 0.42 & 0.43 \\
& ITFormer & TS & 0.33 & 0.37 & 0.29 & 0.31 & 0.37 & 0.29 & 0.36 & 0.36 & 0.37 & 0.40 & 0.35 & 0.35 & 0.35 \\
& ChatTime & TS & -- & -- & -- & -- & -- & -- & -- & -- & -- & -- & -- & -- & -- \\
\midrule
\multirow{1}{*}{Human}
& Experts & -- & 0.67 & 0.71 & 0.59 & 0.67 & 0.75 & 0.75 & 0.80 & 0.66 & 0.66 & 0.72 & 0.63 & 0.67 & 0.58 \\
\bottomrule
\end{tabular}
}
\end{table*}

\begin{table*}[ht]
\centering
\caption{Performance of different models on the \textbf{Match} subset.
$^{cot}$ denotes \textit{thinking} mode.
$^1$ denotes models evaluated without any time series input.
$^2$ denotes ChatTS without built-in statistical computation module.
\texttt{-VL} = Vision-Language.
'TS' stands for Time Series modality, as time-series-specific models introduce a TS encoder.
'--' indicates that the model failed to respond correctly.
\textbf{\underline{Bold underlined}} values indicate the best performance within each category for each metric, and \textbf{bold} values indicate the second-best performance.}
\label{tab:MMTS-Match-performance}
\renewcommand{\arraystretch}{1.2} 
\setlength{\tabcolsep}{4pt}       
\resizebox{\textwidth}{!}{
\begin{tabular}{l|l c|c|c c c c}
\toprule
\textbf{Category} & \textbf{Model Name} & \textbf{Modality} & \textbf{Average} & \textbf{Isomorphic} & \textbf{Robust} & \textbf{Localization} & \textbf{Reverse} \\
\midrule
\multirow{13}{*}{Open-source}
& DeepSeek-V3 & Text & \textbf{\underline{0.65}} & \textbf{\underline{0.90}} & \textbf{\underline{0.79}} & \textbf{\underline{0.57}} & 0.35 \\
& Kimi-K2 & Text & 0.60 & 0.78 & 0.66 & 0.50 & 0.44 \\
& Qwen3-32b$^{cot}$ & Text & 0.60 & 0.80 & 0.64 & 0.42 & \textbf{\underline{0.52}} \\
& Qwen3-32b & Text & 0.50 & 0.72 & 0.58 & 0.36 & 0.35 \\
& Qwen3-8b$^{cot}$ & Text & 0.50 & 0.62 & 0.56 & 0.36 & \textbf{0.46} \\
& Qwen3-8b & Text & 0.42 & 0.56 & 0.48 & 0.31 & 0.32 \\
& Qwen2.5-32b & Text & \textbf{0.62} & \textbf{0.84} & \textbf{0.72} & \textbf{0.53} & 0.41 \\
& Qwen2.5-14b & Text & 0.57 & 0.78 & 0.61 & 0.42 & 0.45 \\
& Qwen2.5-7b & Text & 0.40 & 0.45 & 0.49 & 0.35 & 0.29 \\
& Qwen2.5-7b-VL & Text & 0.30 & 0.31 & 0.36 & 0.27 & 0.28 \\
& Qwen2.5-7b-VL & Vision & 0.28 & 0.30 & 0.31 & 0.28 & 0.25 \\
& Qwen2.5-7b-VL & V+T & 0.34 & 0.40 & 0.41 & 0.27 & 0.30 \\
& Qwen2.5-32b$^{1}$ & Text & 0.25 & 0.25 & 0.25 & 0.25 & 0.25 \\
\midrule
\multirow{11}{*}{Closed-source}
& Claude-3.7-Sonnet & Text & 0.74 & 0.93 & \textbf{0.81} & \textbf{\underline{0.67}} & 0.54 \\
& Claude-Sonnet-4 & Text & 0.71 & 0.85 & 0.79 & \textbf{0.61} & 0.60 \\
& Gemini-2.5-Pro & Text & \textbf{0.79} & \textbf{0.96} & 0.80 & 0.59 & \textbf{0.80} \\
& Gemini-2.5-Flash & Text & 0.44 & 0.63 & 0.57 & 0.28 & 0.26 \\
& GPT-5-High & Text & \textbf{\underline{0.81}} & \textbf{\underline{0.98}} & \textbf{0.81} & 0.60 & \textbf{\underline{0.86}} \\
& GPT-5-Minimal & Text & 0.57 & 0.84 & 0.67 & 0.53 & 0.26 \\
& GPT-4.1 & Text & 0.67 & 0.89 & \textbf{\underline{0.82}} & 0.55 & 0.40 \\
& GPT-4.1-Mini & Text & 0.63 & 0.90 & 0.78 & 0.44 & 0.40 \\
& GPT-4o & Text & 0.50 & 0.68 & 0.60 & 0.41 & 0.34 \\
& GPT-4o & Vision & 0.45 & 0.56 & 0.50 & 0.34 & 0.38 \\
& GPT-4o & V+T & 0.55 & 0.79 & 0.64 & 0.38 & 0.38 \\
\midrule
\multirow{4}{*}{Time-series}
& ChatTS & TS & 0.37 & 0.47 & 0.40 & 0.24 & 0.36 \\
& ChatTS$^{2}$ & TS & 0.32 & 0.46 & 0.41 & 0.22 & 0.20 \\
& ITFormer & TS & 0.24 & 0.16 & 0.25 & 0.25 & 0.29 \\
& ChatTime & TS & -- & -- & -- & -- & -- \\
\bottomrule
\end{tabular}
}
\end{table*}


\begin{table*}[ht]
\centering
\caption{Performance of different models on the \textbf{Align} subset.
$^{cot}$ denotes \textit{thinking} mode.
$^1$ denotes models evaluated without any time series input.
$^2$ denotes ChatTS without built-in statistical computation module.
\texttt{-VL} = Vision-Language.
'TS' stands for Time Series modality, as time-series-specific models introduce a TS encoder.
'--' indicates that the model failed to respond correctly.
\textbf{\underline{Bold underlined}} values indicate the best performance within each category for each metric, and \textbf{bold} values indicate the second-best performance.}
\label{tab:MMTS-Align-performance-2}
\renewcommand{\arraystretch}{1.2} 
\setlength{\tabcolsep}{4pt}       
{
\begin{tabular}{l|l c|c|c c}
\toprule
\textbf{Category} & \textbf{Model Name} & \textbf{Modality} & \textbf{Average} & \textbf{TS$\to$Sem} & \textbf{Sem$\to$TS} \\
\midrule
\multirow{13}{*}{Open-source}
& DeepSeek-V3 & Text & \textbf{0.94} & \textbf{\underline{0.95}} & \textbf{0.94} \\
& Kimi-K2 & Text & \textbf{\underline{0.95}} & \textbf{0.94} & \textbf{\underline{0.96}} \\
& Qwen2.5-32b & Text & 0.93 & 0.92 & \textbf{0.94} \\
& Qwen2.5-14b & Text & 0.88 & 0.87 & 0.88 \\
& Qwen2.5-7b & Text & 0.69 & 0.68 & 0.71 \\
& Qwen3-32b$^{cot}$ & Text & 0.89 & 0.92 & 0.86 \\
& Qwen3-32b & Text & 0.87 & 0.86 & 0.88 \\
& Qwen3-8b$^{cot}$ & Text & 0.86 & 0.88 & 0.83 \\
& Qwen3-8b & Text & 0.79 & 0.74 & 0.84 \\
& Qwen2.5-7b-VL & Text & 0.64 & 0.68 & 0.59 \\
& Qwen2.5-7b-VL & Vision & 0.60 & 0.61 & 0.60 \\
& Qwen2.5-7b-VL & V+T & 0.73 & 0.78 & 0.67 \\
& Qwen2.5-32b$^{1}$ & Text & 0.27 & 0.29 & 0.26 \\
\midrule
\multirow{8}{*}{Closed-source}
& Claude-3.7-Sonnet & Text & 0.97 & 0.97 & \textbf{0.98} \\
& Claude-Sonnet-4 & Text & \textbf{0.98} & \textbf{0.98} & \textbf{\underline{0.99}} \\
& Gemini-2.5-Pro & Text & 0.97 & 0.97 & \textbf{\underline{0.99}} \\
& Gemini-2.5-Flash & Text & 0.94 & 0.94 & 0.95 \\
& GPT-5-High & Text & \textbf{\underline{0.99}} & \textbf{\underline{0.99}} & \textbf{\underline{0.99}} \\
& GPT-5-Minimal & Text & 0.97 & 0.97 & \textbf{0.98} \\
& GPT-4o & Text & 0.96 & 0.96 & 0.97 \\
& GPT-4o-Mini & Text & 0.86 & 0.82 & 0.90 \\
\midrule
\multirow{4}{*}{Time-series}
& ChatTS & TS & 0.80 & 0.68 & 0.91 \\
& ChatTS$^{2}$ & TS & 0.45 & 0.49 & 0.42 \\
& ITFormer & TS & 0.29 & 0.32 & 0.26 \\
& ChatTime & TS & -- & -- & -- \\
\bottomrule
\end{tabular}
}
\end{table*}

\begin{table*}[ht]
\centering
\begin{minipage}[t]{0.48\textwidth}
\centering
\caption{Performance of the ChatTS model on the \textbf{Match} subset across different time series length ranges. \textit{Total} corresponds to the range [13,504], and "–" indicates that no questions fall into the given length range for that task.}
\label{chatts_on_different_ts_length}
\begin{tabular}{lcccc}
\toprule
Length Range & Isomorphic & Robust & Localization & Reverse \\
\midrule
Total       & 0.47 & 0.40 & 0.24 & 0.36 \\
{[64, 1024]}  & \textbf{0.53} & \textbf{0.57} & 0.23 & \textbf{0.44} \\
{[256, 512]}  & --   & --   & \textbf{0.36} & --   \\
\bottomrule
\end{tabular}
\end{minipage}
\hfill
\begin{minipage}[t]{0.48\textwidth}
\centering
\caption{Token cost for a single evaluation on the InWild subset (1,084 samples).}
\label{tab:token_cost_inwild}
\begin{tabular}{lccc}
\toprule
\textbf{Model} & \textbf{Input/Output Tokens} & \textbf{Price Cost} \\
\midrule
Qwen2.5-32B    & $\approx$6M / $\approx$20k  & 0 \\
DeepSeek-V3     & $\approx$5M / $\approx$250k & $\approx$\$1.62 \\
GPT-4o          & $\approx$5M / $\approx$400k & $\approx$\$16.50 \\
Claude-Sonnet-4 & $\approx$5M / $\approx$600k & $\approx$\$24.00 \\
Gemini-2.5-Pro  & $\approx$6M / $\approx$500k & $\approx$\$12.50 \\
\bottomrule
\end{tabular}
\end{minipage}
\end{table*}



\clearpage
\section{Statistical Robustness Analysis}
\label{sec: robustness}


To assess the testing stability, robustness, and validity of \textsc{\textsc{MMTS-Bench}}, we conducted comprehensive statistical evaluations and bias analyses: \textbf{Bootstrap Confidence Interval}, \textbf{Iterative Subsampling Analysis}, and \textbf{Assessment of Dataset Artifacts}. These experiments evaluate the dataset's testing stability, the adequacy of its scale, and its resistance to spurious correlations, respectively.

\subsection{Bootstrap Confidence Interval}
\label{sec: robustness1}

To evaluate the reliability of model performance and quantify its uncertainty, we adopted the non-parametric bootstrap method. Specifically, the experimental setup is as follows: Given a test set of size $D$, we generated $N=1000$ bootstrap samples, also of size $D$, through sampling with replacement.

We selected a few representative models (covering close-source, open-source LLMs and TS-LLMs) for the \textbf{InWild} subset, the \texttt{choice} split of the \textbf{Base} subset, and the entire \textsc{\textsc{MMTS-Bench}} dataset. We then calculated their respective accuracy scores to obtain an empirical distribution for this metric. Based on this distribution, we report the mean accuracy, standard deviation (std), and the 95\% confidence interval (CI).

\begin{table}[h!]
\centering
\caption{Bootstrap confidence interval of different models on the MMTS-InWild dataset.}
\label{tab:bootstrap_inwild}
\begin{tabular}{l|cccc}
\toprule
\textbf{Models} & \textbf{Mean} & \textbf{Std} & \textbf{CI Low} & \textbf{CI High} \\
\midrule
Gemini-1.5-Pro (text) & 0.6827 & 0.0142 & 0.6541 & 0.7103 \\
Gemini-2.5-Pro (vision) & 0.7262 & 0.0138 & 0.6983 & 0.7537 \\
GPT-4o (text) & 0.6128 & 0.0148 & 0.5830 & 0.6421 \\
DeepSeekV3 (text) & 0.6218 & 0.0150 & 0.5932 & 0.6504 \\
Qwen2.5-32b (text) & 0.5379 & 0.0156 & 0.5083 & 0.5683 \\
\bottomrule
\end{tabular}
\end{table}

\begin{table}[h!]
\centering
\caption{Bootstrap confidence interval of different models on the MMTS-Base's choice split.}
\label{tab:bootstrap_base}
\begin{tabular}{l|cccc}
\toprule
\textbf{Models} & \textbf{Mean} & \textbf{Std} & \textbf{CI Low} & \textbf{CI High} \\
\midrule
Gemini-1.5-Pro (text) & 0.6340 & 0.0205 & 0.5933 & 0.6726 \\
Claude-3.7-Sonnet (text) & 0.5967 & 0.0207 & 0.5581 & 0.6391 \\
GPT-4o (text) & 0.5609 & 0.0210 & 0.5211 & 0.6021 \\
DeepSeekV3 (text) & 0.5433 & 0.0205 & 0.5053 & 0.5845 \\
Qwen2.5-32b (text) & 0.4727 & 0.0218 & 0.4331 & 0.5158 \\
\bottomrule
\end{tabular}
\end{table}

\begin{table}[h!]
\centering
\caption{Bootstrap confidence interval of different models on the complete \textsc{MMTS-Bench} dataset.}
\label{tab:bootstrap_full}
\begin{tabular}{l|cccc}
\toprule
\textbf{Models} & \textbf{Mean} & \textbf{Std} & \textbf{CI Low} & \textbf{CI High} \\
\midrule
Gemini-1.5-Pro (text) & 0.7235 & 0.0094 & 0.7059 & 0.7422 \\
GPT-4o (text) & 0.6059 & 0.0103 & 0.5864 & 0.6265 \\
DeepSeekV3 (text) & 0.6366 & 0.0100 & 0.6165 & 0.6562 \\
Qwen2.5-32b (text) & 0.5789 & 0.0104 & 0.5580 & 0.5986 \\
\bottomrule
\end{tabular}
\end{table}

\begin{figure*}[htbp]
    \centering
    \begin{subfigure}{\textwidth}
        \centering
        \includegraphics[width=\textwidth, height=0.27\textheight, keepaspectratio]{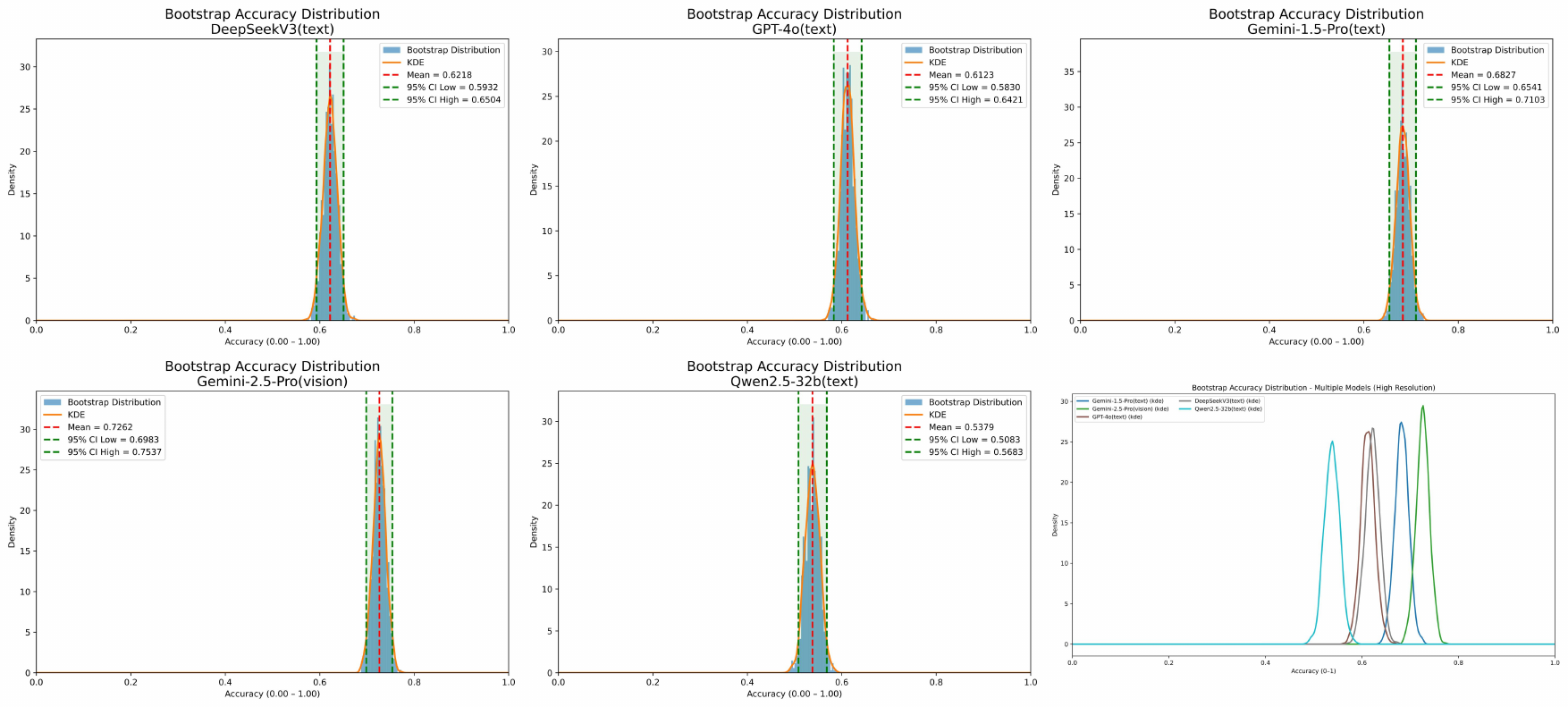}
        \caption{MMTS-InWild}
    \end{subfigure}\\[1em]
    \begin{subfigure}{\textwidth}
        \centering
        \includegraphics[width=\textwidth, height=0.27\textheight, keepaspectratio]{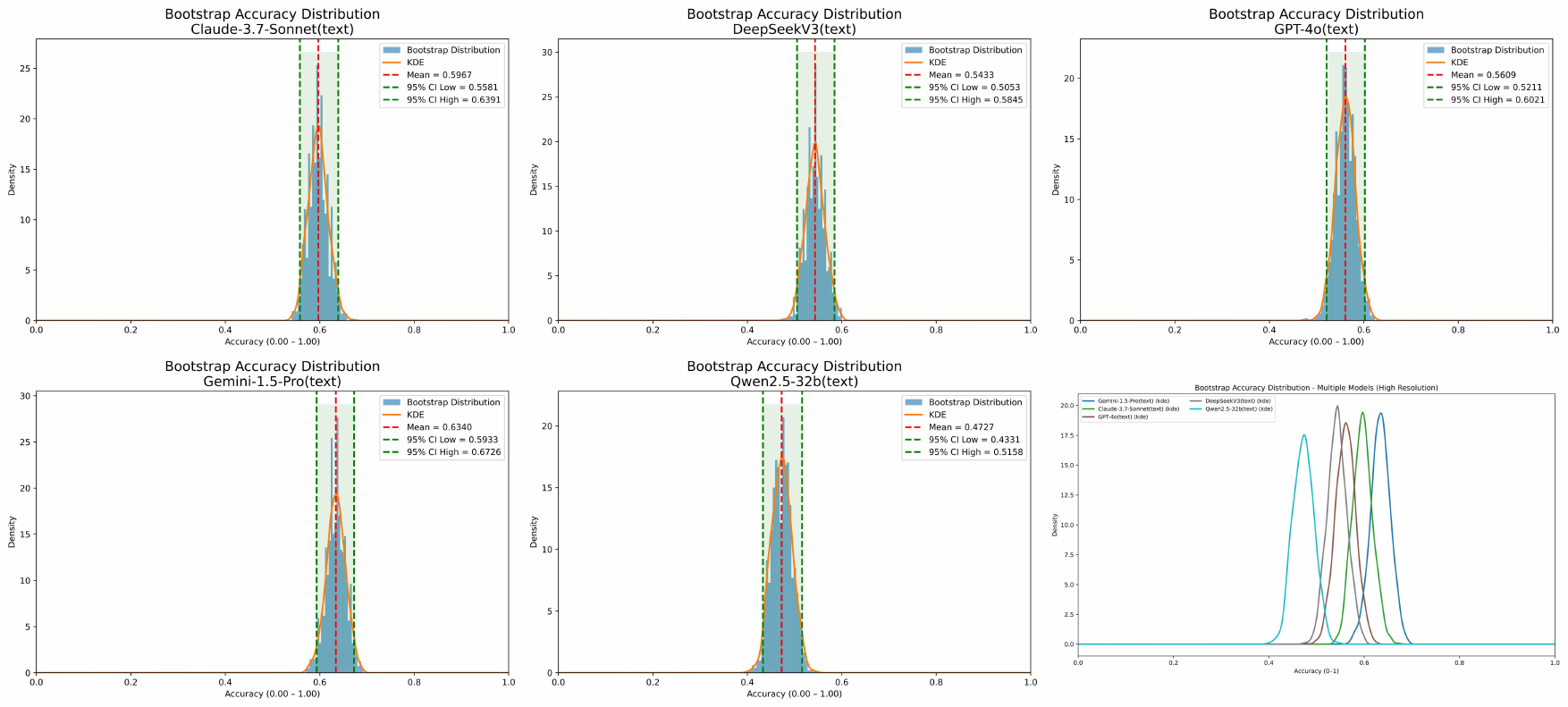}
        \caption{MMTS-Base (choice split)}
    \end{subfigure}\\[1em]
    \begin{subfigure}{\textwidth}
        \centering
        \includegraphics[width=\textwidth, height=0.27\textheight, keepaspectratio]{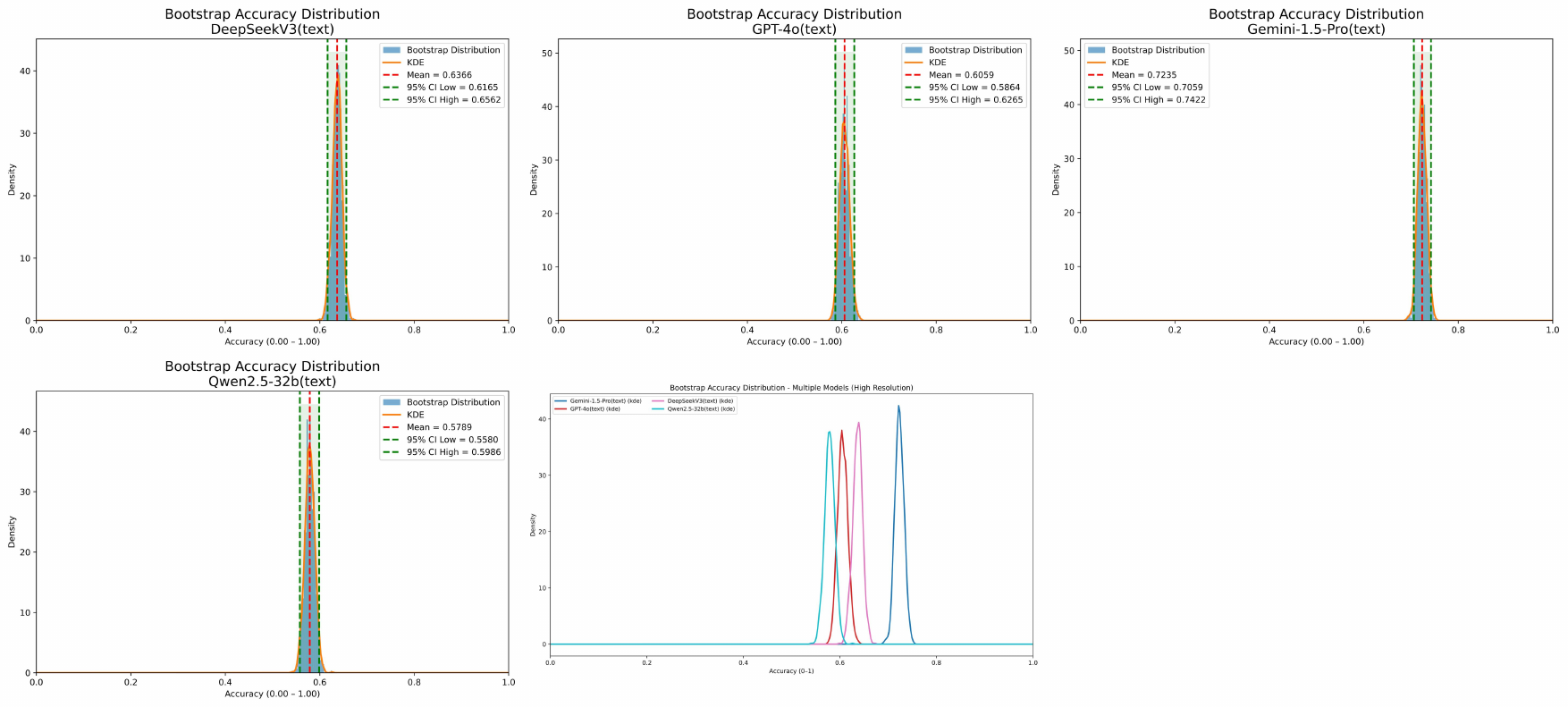}
        \caption{\textsc{MMTS-Bench}}
    \end{subfigure}
    \caption{Visualization of the accuracy distribution and confidence intervals for different representative models.}
    \label{fig:bootstrap}
\end{figure*}

    




The experimental results (Table ~\ref{tab:bootstrap_inwild}, ~\ref{tab:bootstrap_base}, ~\ref{tab:bootstrap_full} and Figure ~\ref{fig:bootstrap}) indicate that the performance evaluations across all models exhibit low statistical dispersion, both on the full dataset and the subsets. Specifically, the bootstrapping experiment shows that the standard deviation (std) ranges only between $0.01 \sim 0.02$, and the width of the 95\% CI remains within a narrow range (approximately $0.5 \sim 0.8$ percentage points).

These tight error bounds strongly confirm the statistical robustness of the \textsc{\textsc{MMTS-Bench}}. It demonstrates that the benchmark is insensitive to data sampling variance and can provide stable and reproducible evaluation results for cross-model and cross-capability comparisons. Furthermore, to further suppress the stochastic noise from model generations, we also introduced a mechanism of multiple sampling and majority voting during the evaluation, thereby establishing a more robust performance baseline.

\subsection{Iterative Subsampling Analysis}
\label{sec: robustness2}

To investigate the relationship between evaluation stability and dataset scale, and to estimate the minimal sample size required to yield robust results, we conducted an Iterative Subsampling Analysis focusing on the representative model, \texttt{Gemini-2.5-Pro}, with text input.

The specific experimental setup is as follows: For a given dataset size $D$, we set the subsampling size $S$ as a variable that progressively increases from an initial value up to $D$, with an increment step of $T=20$. At each fixed size $S$, we perform $N=50$ independent repetitions of sampling, and calculate the mean, standard deviation (std), and coefficient of variation (CV) of the model's performance.

We use the coefficient of variation ($CV = \sigma / \mu$) as the core metric to measure evaluation stability. The dataset size $S$ is deemed to possess sufficient statistical stability when the CV curve, as $S$ increases, shows a descending trend and falls below a pre-set convergence threshold of $\tau = 0.02$. \footnote{We empirically set the convergence threshold to $\tau = 0.02$, which requires the standard deviation of the evaluation score to be controlled within $2\%$ of the mean. For a typical model accuracy range ($50\% \sim 80\%$), this means the measurement error is limited to an absolute range of approximately $1\% \sim 1.6\%$. This strict stability constraint is crucial for suppressing ``ranking flips'' caused by sampling variance, ensuring the benchmark can reliably distinguish between models with slight performance differences.} This experiment covered the entire \textsc{MMTS-Bench} and its four subsets.

\begin{table}[htbp]
    \centering
    \caption{Comparison of the minimum required sample size for stable assessment versus the actual sample size in the subsampling analysis experiment, along with the model's mean, standard deviation, and coefficient of variation under the actual sample size across four data subsets and the entire \textsc{MMTS-Bench} dataset.}
    \label{tab:subsampling}
    \resizebox{\columnwidth}{!}{%
    \begin{tabular}{l|ccccc}
        \toprule
        \textbf{Dataset} & \textbf{Full} & \textbf{Min} & \textbf{Mean} & \textbf{Std} & \textbf{CV} \\
        \midrule
        Align & 240 & 60 & 0.9813 & 0.0100 & 0.0041 \\
        Base (choice) & 568 & 400 & 0.6357 & 0.0203 & 0.0038 \\
        Match & 400 & 260 & 0.7795 & 0.0199 & 0.0047 \\
        InWild & 1084 & 600 & 0.6811 & 0.0130 & 0.0011 \\
        Full & 2292 & 600 & 0.7199 & 0.0093 & 0.0010 \\
        \bottomrule
    \end{tabular}%
    }
\end{table}


\begin{figure*}[t]
\begin{center}
\includegraphics[width=\textwidth]{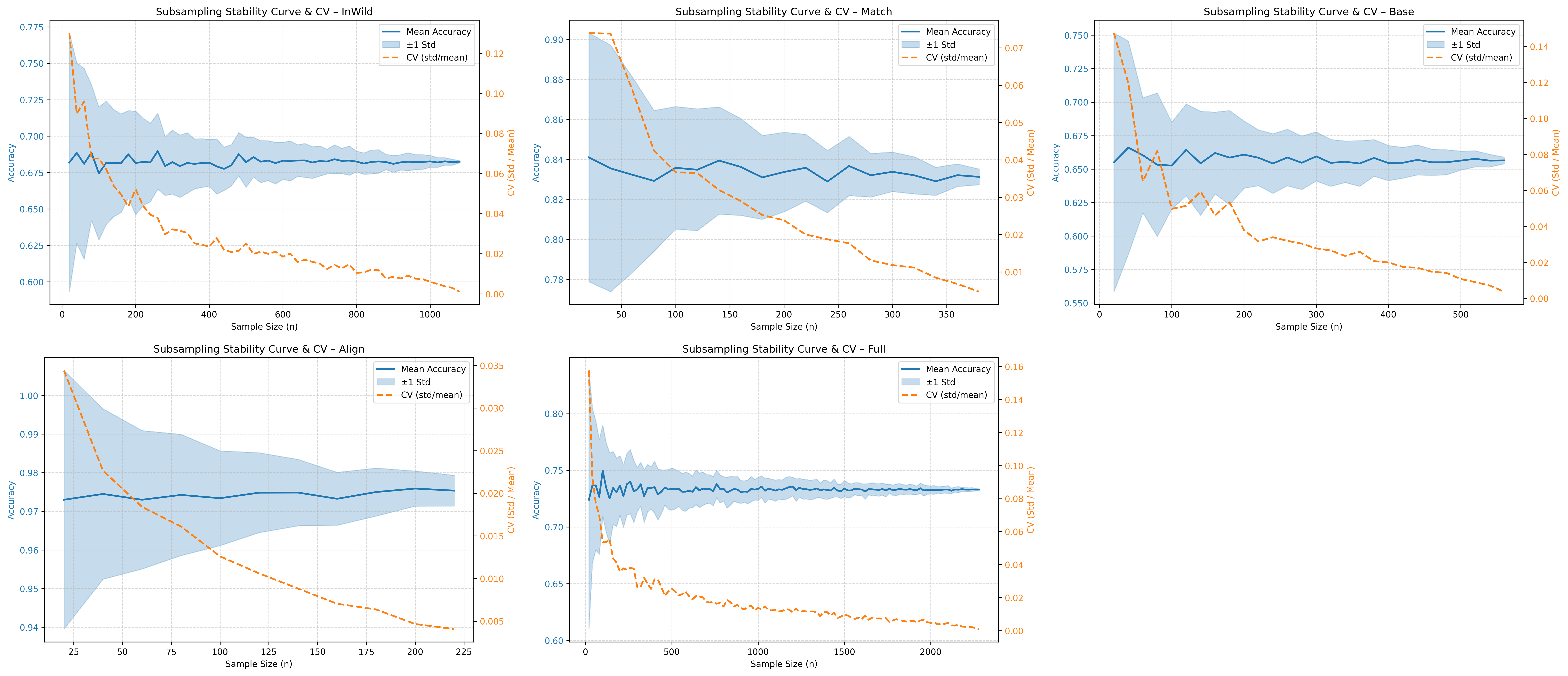}
\end{center}
\caption{Trends of model accuracy metrics (mean, standard deviation, and coefficient of variation) with varying subsampling size on four subsets and the entire \textsc{MMTS-Bench}.}
\label{fig:subsampling}
\end{figure*}

The experimental results (Table ~\ref{tab:subsampling} and Figure ~\ref{fig:subsampling}) demonstrate that the current data scale of \textsc{MMTS-Bench} provides an ample safety margin for evaluation stability. Specifically, the actual sample size of each subset significantly exceeds the minimum number of samples required to reach the convergence threshold (approximately $1.42 \sim 4$ times the required minimum), and the lowest CV has dropped to the $10^{-3}$ magnitude. This outcome confirms that we have not only ensured  high confidence in the evaluation results at the current scale but have also achieved a good balance between statistical robustness and evaluation efficiency (computational and time costs).

\subsection{Assessment of Dataset Artifacts and Shortcut Learning}
\label{sec:artifacts}

A substantial body of research warns against benchmark performance driven by spurious correlations or explicit features rather than genuine reasoning \citep{geirhos2020shortcut, gururangan2018annotation}. To ensure \textsc{MMTS-Bench} evaluates robust time-series reasoning rather than relying on dataset artifacts, we analyzed the dependency of model performance on explicit surface-level attributes. Specifically, we examined the correlation between TSQA accuracy and three explicit factors: sequence length ($L$), variable count ($V$), and question text length ($T$) on the InWild subset. We evaluated three representative models: GPT-4o, Qwen2.5-32B, and ChatTS\citep{xie2024chatts}. We introduce three metrics to quantify these dependencies: \textbf{(1) Correlation ($r_L, r_T$).} The Pearson correlation coefficient between accuracy and the logarithm of sequence length ($r_L$) or question text length ($r_T$). A value close to 0 indicates no linear dependency. \textbf{(2) Length Sensitivity ($\Delta_{\text{long}}$).} The difference in mean accuracy between the samples in the longest quartile ($\ge 75$th percentile) and the shortest quartile ($\le 25$th percentile). \textbf{(3) Dimensionality Gap ($\Delta_{\text{dim}}$).} The difference in mean accuracy between multivariate and univariate samples.

As presented in Table~\ref{tab:artifact_analysis}, the results reveal minimal dependence on these artifacts. The correlation with sequence length is negligible across all models ($|r_L| < 0.08$), and the accuracy gap between extreme lengths ($\Delta_{\text{long}}$) remains within a narrow range (approx. $0.05 \sim 0.08$), showing no consistent bias towards short or long sequences. Similarly, the performance gap between univariate and multivariate series is marginal ($|\Delta_{\text{dim}}| < 0.04$), and question length shows only a weak effect ($|r_T| \approx 0.15$). These findings confirm that \textsc{MMTS-Bench} performance is not trivially predictable by simple metadata features.

\begin{table}[h]
\centering
\caption{Analysis of potential dataset artifacts and shortcut learning on the InWild subset.}
\label{tab:artifact_analysis}
\begin{tabular}{lcccc}
\toprule
\textbf{Model Name} & \textbf{$r_L$} & \textbf{$\Delta_{\text{long}}$} & \textbf{$\Delta_{\text{dim}}$} & \textbf{$r_T$} \\
\midrule
GPT-4o & -0.0693 & 0.0824 & -0.0101 & -0.1451 \\
Qwen2.5-32B & 0.0547 & -0.0525 & 0.0353 & -0.0264 \\
ChatTS & 0.0727 & -0.0502 & -0.0163 & -0.0862 \\
\bottomrule
\end{tabular}
\end{table}

\subsection{Conclusion}

We conducted three systematic analyses—Bootstrap Confidence Interval, Iterative Subsampling Analysis, and Assessment of Dataset Artifacts—which collectively demonstrate that \textsc{MMTS-Bench} possesses high statistical robustness, an efficient scale, and validity against shortcut learning. The results confirm that model performance on \textsc{MMTS-Bench} is driven by genuine time-series understanding rather than explicit surface-level features (e.g., sequence length or dimensionality), ensuring that the evaluation results are stable, reliable, and trustworthy.


\section{Reproducibility Statement}
\label{sec:reproducibility statement}
We elaborate on the implementation details of our benchmark construction and experimental setup in this paper and the Appendix. To facilitate end-to-end reproduction, we release an anonymized repository containing all data and code at \url{https://anonymous.4open.science/r/MMTS-BENCH-BEF7/}. We will maintain the anonymized repository for the duration of the review and, upon acceptance, migrate to a public repository and archive a snapshot to support long-term availability.









\end{document}